\documentclass[twocolumn,groupeaddress,superscriptaddress,showpacs,cite]{revtex4-1}
\usepackage{epsfig,pslatex,latexsym,times,amssymb,amsmath,graphicx,bm,revsymb}
\pdfoutput=1

\begin{document}
\author{M. Prada}
\email{mprada@physnet.uni-hamburg.de}
%\affiliation{I. Institut f\"ur Theoretische Physik, Universit\"at Hamburg, Jungiusstr. 9, 20355 Hamburg, Germany}
%
\author{D. Pfannkuche}
\affiliation{I. Institut f\"ur Theoretische Physik, Universit\"at Hamburg, Jungiusstr. 9, 20355 Hamburg, Germany}
\date{\today}
%
%\title{Anisotropy of spin coherence in high mobility quantum wells with arbitrary magnetic fields} 
\title{Anisotropic decoherence in quantum wells with arbitrary magnetic fields: Interplay of the spin-orbit coupling terms} 
\date{\today}
\newcommand*{\ud}{\mathrm{d}}
\newcommand*{\ue}{\mathrm{e}}
\begin{abstract}
{ 
We present a theoretical study of the anisotropy of the spin relaxation and decoherence  % derive an expression for the anisotropy of the spin relaxation and decoherence 
in typical quantum wells with an arbitrary magnetic field.  %in the presence of a 
In such systems, the orientation of the magnetic field relative to the main crystallographic 
directions is crucial, % as it determines the interplay between spin orbit coupling and spin precession 
owing to the lack of spin-rotation symmetry.  
%In such systems, the relative orientation of two coordinate systems is crucial: 
%the crystallographic axis, where electronic motion results in spin-orbit coupling 
%and the magnetic axis, determining the quantization direction of the spin.
%The natural choice of the Hilbert space for the perturbation is thus the former, 
%whereas the latter marks the perturbation
For typical high mobility samples, relaxation anisotropies %in the motional narrowing limit 
owing to the interplay of % enhanced in the presence of both  %due to the interplay of the
Rashba and Dresselhaus spin orbit coupling are calculated. 
%This suggest a scheme for the experimental determination of the relative strength of Rashba and 
%Dresselhaus couplings using ESR. % can thus be determined by ESR. 
We also include  the effect of the cubic-in-momentum terms. 
Although commonly ignored in literature, the latter were experimentally evidenced by 
the observation of strong anisotropy in spin decoherence measurements by 
different experimental groups and has long remained unexplained.
This work suggests a method to determine the relative strength of spin-orbit coupling terms by angular resolution of decoherence in ESR experiments. 

%Truitt {\it et al.} \cite{truitt}. 
}
\end{abstract}

%\pacs{76.60.Es,71.70.Ej, 68.65.Fg,73.21.Fg}

\maketitle
%%%%%%%%%%%%%%%%%%%%%%%%%%%%%%%%%%%%%%%%%%%%%%%%%%%%%%%%%%%%%
%%%%%%%%%%%%%%%%%%%%%%%%%%%%%%%%%%%%%%%%%%%%%%%%%%%%%%%%%%%%%
\section{Introduction}
Spin relaxation processes in semiconductors continue to attract attention in connection with 
a large number of spintronic applications \cite{reviewSi,SiComp,rafa,datadas,zutic}, 
in which silicon appears to be a very suitable material due to long decoherence times 
\cite{tahan,kawakami,tyryshkin, tyryshkin_prl} and high gate fidelity \cite{wong, rohling,veldhorst}.
%g-factor tuning has been achieved in Si quantum wells (QWs) \cite{wilamowski_prl}, 
Electron spin resonance (ESR) is a promising technique to manipulate spins directly 
in high-mobility quantum well heterostructures (QW) \cite{loss,sanada,koppens}. 
In these systems, the main source of decoherence is typically spin-orbit coupling (SOC)
due to bulk (BIA), structure (SIA) or interface inversion asymmetry (IIA)  \cite{wrinkler,sherman, vervoort}. %prada_njop,nestoklon}. 
The interplay of the different SOC contributions may result in strong anisotropy of spin  relaxation and decoherence, 
giving a hint on their relative strength. 
%However, the strengths of the SOC terms are difficult to measure independently, 
%and a full understanding of their interplay is crucial for spintronic based applications. % to making such devices. 
%Additionally, in confined systems such as quantum dots, some effects of the linear 
%SO terms are suppressed \cite{averin}.

\begin{figure}[!hbt]
\centering\includegraphics[angle=0, width = 0.35\textwidth]{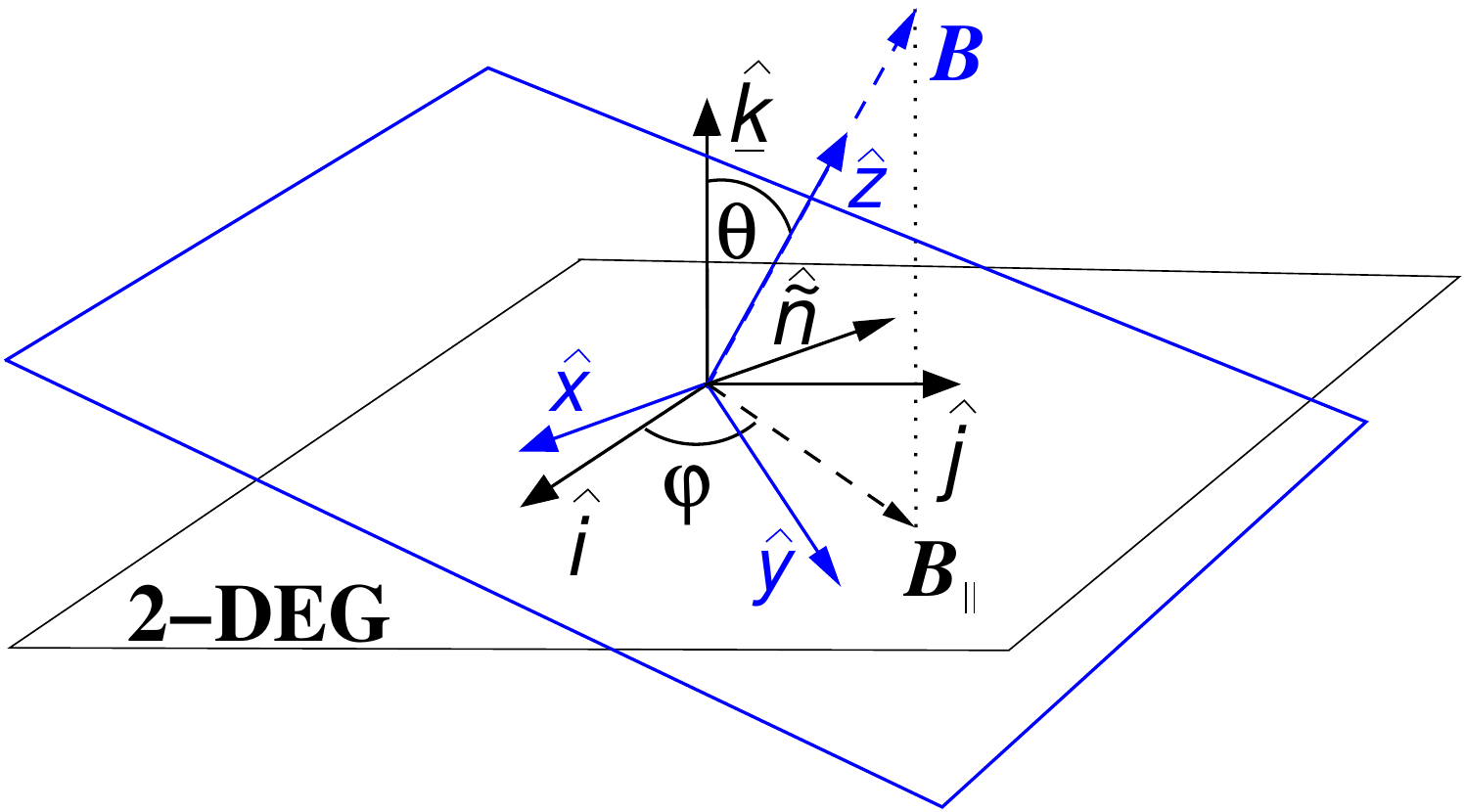}
\caption{\it \footnotesize The main quantization axis $\hat z$ is given by $\bm B$, 
determining the magnetic coordinate system (blue), with $\hat x, \hat y$ being arbitrary directions in the plane perpendicular to $\hat z$. 
%We choose the projection of $\hat x$ and the projection of $\bm B$ in the $\hat i-\hat j$ plane to be parallel, for simplicity. 
The crystallographic coordinate system typically define the electronic motion ($\hat i$, $\hat j$) and the growth direction ($\hat{k}$) (black). 
Any (pseudo-) vector can be expressed in either coordinate system, with its components 
being related by the Euler-Rodrigues rotation matrix, $R_{\hat n}(\theta)$.
}
\label{fig1}
\end{figure}

In this work we describe decoherence and relaxation processes in general 2-DEG ESR experiments, quantifying 
the anisotropies in terms of the relative direction of external fields and main crystallographic directions. 
The appropriate choice of coordinate system is crucial in the calculations of spin-related observables, owing to the 
anisotropy that follows the interplay of the different contributions of SOC. 
Scattering relaxation events and SOC are defined within the 2-DEG perpendicular to the growth direction $\hat k$,  
determining the crystallographic coordinate system $\{\hat i, \hat j, \hat k\}$ (black arrows of Fig. \ref{fig1}),   
whereas the quantization axis is naturally determined by the direction of the magnetic field $\hat z$. 
%The natural choice of the quantization axis  suggests thus expressing the crystallographic coordinate system 
%$\{\hat i, \hat j, \hat {k} \}$ in terms of the magnetic one, $\{\hat x, \hat y, \hat z\}$.
%Defining $\theta, \phi$ as the azimuthal and polar angles, respectively. 
The magnetic coordinate system  $\{\hat x, \hat y, \hat z\}$ (blue arrows of Fig. \ref{fig1}), 
%and the corresponding polar and azimuthal angle $\theta, \varphi$ that relates both axial systems. 
is  related to the crystallographic one by the Euler-Rodrigues formula, consisting of a rotation along $\hat n$ 
by the polar angle $\theta$ (see Fig. \ref{fig1}).
During an ESR experiment, the magnetic field direction is changed relative to the
crystallographic axis, such that $\theta$ varies while the azimuth $\varphi$ remains constant. 
For $\theta =0$, spin-rotation symmetry  SU(2) is %reduced to U(1) in the presence of a single source of SOC or 
completely broken (reduced to U(1)), if both BIA and SIA (only BIA or only SIA) SOC are present \cite{lyanda}.  
As $\theta$ is varied,  spin-rotation symmetry is completely broken, and a measurement of the anisotropy 
in spin observables is then expected to be captured in terms of the azimuthal angle. 
As a result, the relative strength of the different SOC contributions can be experimentally determined in 
angle-resolved measurements. 

The  aim of this work is to achieve a complete description of D'yakonov-Perel'  decoherence and relaxation processes as a function of 
the main angles, capturing the effects of the anisotropy that results from the interplay of BIA and SIA. 
We find that the choice of the magnetic coordinate system possesses clear advantages, %as the time dependency factorizes, 
rendering Redfield-type  equations of motion with analytical solutions.  
We  also include the effects of the commonly neglected cubic-in-momentum terms, which may become important in typical experiments. 
Our theoretical model finds an excellent agreement with experimental data on high-mobility QWs, where the existence of strong 
anisotropies have long remained unexplained. 

This article is organized as follows:  
in  Section \ref{sec:methods}, we  describe spin relaxation and decoherence processes. 
First, (\ref{sec:methodsA}) we define the SOC in the magnetic coordinate system, then we write the equations of motion within 
the Redfield approach (\ref{sec:methodsB}), and finally, we obtain 
an expression for $T_1$ and $T_2$ (\ref{sec:methodsC}), 
first for a perpendicular magnetic field with linear SOC terms, then for magnetic fields with  arbitrary direction and last, 
including the effects of cubic in momentum SOC terms. 
Section \ref{sec:results} is devoted to results, from a theoretical perspective (\ref{sec:resultsA}) and 
in connection with  experiments (\ref{sec:resultsB}). 
We finally give some concluding remarks in section \ref{sec:conclusion}. 

\section{Methods}
\label{sec:methods}
\subsection{SOC terms in magnetic coordinates}
\label{sec:methodsA}
%We calculate decoherence and relaxation in a typical ESR experiment performed on 
%high mobility two-dimensional electron system (2DES):
%We consider the ESR experiment performed by J. Truitt \cite{truitt}, using a
%Si$_{1-x}$Ge$_{x}$/Si/Si$_{1-x}$Ge$_{x}$ quantum well of known mobility
%(and thus, momentum relaxation time, $\tau_k$). \
We consider a general 2-DEG sample with an external magnetic field $\bm B$  %a varying polar angle $\theta$ 
whose direction is defined by the polar angle $\theta$ 
and %with the growth axis $\hat {k}$, whereas 
the azimuth
$\varphi$. The latter corresponds to angle of the main crystallographic direction $\hat i$ and  the 
projection of the magnetic field onto the 2-DEG (see Fig. \ref{fig1}). 
The direction of the magnetic field $\bm B = B\sigma_z$ determines the  magnetic coordinate system, $\hat x, \hat y, \hat z$ 
and the quantization of the spin, $\sigma_z$, 
\begin{equation}
 H_0 = -\frac{1}{2}g\mu_B B \sigma_z.
\label{eq:H0}
\end{equation}
%with the growth direction of the sample (see Fig. 1). 
%The natural choice of the Hilbert space for the unperturbed 
%Hamiltonian is thus spanned by the Pauli matrices $\{\sigma_\alpha\}$, $\alpha=x,y,z$, where 
%$\sigma_z$ is  {\it always} parallel to $\bm B$, 
%and $\sigma_{x,y}$ have arbitrary directions in the plane perpendicular to $\hat z$, 
Here,  $\mu_B$ is the Bohr magneton and $g$ denotes the $g$-factor, giving a Larmor frequency 
of precession $\omega_l \simeq g\mu_BB/\hbar$. 

Rashba- (SIA) and Dresselhaus- (BIA) type coupling terms appear in the Hamiltonian,   %give rise in the spin Hamiltonian, 
owing to the motion of electrons confined to the 2DEG with broken inversion symmetry.  %with a momentum $\bm { k}$ in the confined 2-DEG. 
Note that BIA terms are not limited to systems without bulk inversion (as GaAs, for instance) but also 
in two-dimensional centrosymmetric-based materials such as Si/SiGe, owing to interface-inversion asymmetry \cite{nestoklon,ganichev2}.  
Moreover, the strength of the BIA- related coupling parameter has been determined for the latter, %tipical Si-SiGe heterostructures, 
yielding a larger value than the SIA coupling \cite{prada_njop}.  
To describe SOC, it is customary to choose the Hilbert space spanned by the Pauli matrices $\{\tilde \sigma_l\}$ 
along the crystallographic directions, $l=i,j,k$, in which the SOC contributions to lowest order in momentum reads:
%with $\hat {\b{k}}$, along the confining direction: % $\{\hat i, \hat j, \hat k\}$. 
\begin{eqnarray}
\label{eq_h}
H_{\rm l} &=& 
%\alpha \left( \vec\sigma\times\vec k \right)\cdot\hat k +
%\gamma_{\rm D} \left[
%\sigma_i k_i(k_j^2-k_k^2) + {\rm cyclic \  permutations}%\sigma_j k_j(k_k^2-k_i^2) +\sigma_k k_k(k_i^2-k_j^2) 
%\right]\nonumber\\
\alpha \left( \bm {\tilde\sigma}\times\bm{\tilde k} \right)\cdot\hat {k} + 
\beta (\tilde\sigma_i \tilde k_i-\tilde \sigma_j \tilde k_j).  %+ \beta_3 (\tilde \sigma_j \tilde k_j \tilde k_i^2 -\tilde \sigma_i \tilde k_i \tilde k_j^2) 
%+ \nonumber \\&+ &
%\alpha_3 (\tilde \sigma_i \tilde k_j \tilde k_i^2 -\tilde \sigma_j \tilde k_i \tilde k_j^2)
%.
%\gamma \left[\tilde \sigma_i k_i(k_j^2-k_k^2) + {\rm cyclic \ permutations}\right], 
\end{eqnarray}%
Here  $\alpha$, $\beta$ %=\gamma_{\rm D} \langle(k^2_j-k^2_k)\rangle = \gamma_{\rm D} \langle(k^2_i-k^2_k)\rangle$ %($\alpha_3$, $\beta_3$) 
determine the strength of the Rashba and Dresselhaus  coupling to lowest order in $k$. Symbol $^\sim$ over $o$ ($\tilde o$)
denotes  that $o$ is expressed in the crystallographic coordinate system.  
\newline Higher order in momentum terms may be added \cite{cubic}, 
\[
H_{\rm c} =
\beta_3 (\tilde \sigma_j \tilde k_j \tilde k_i^2 -\tilde \sigma_i \tilde k_i \tilde k_j^2) + \alpha_3 (\tilde \sigma_i \tilde k_j \tilde k_i^2 -\tilde \sigma_j \tilde k_i \tilde k_j^2), 
\]
with $\alpha_3$, $\beta_3$ being the cubic Rashba and Dresselhaus coupling to third order in momentum. 
Bearing in mind that $\beta$ is proportional to $\langle k_k^2\rangle \propto l^{-1}$ with $l$ being the well width, 
we note that the cubic terms may become important with respect to the linear Dresselhaus one  as 
$l$ increases. 
The cubic terms are also enhanced as the carrier density increases, bearing  a larger in-plane $k_i,k_j$ \cite{wilamowski}. 
%Two general vectors $\bm {\tilde v}$ and $\bm v$
%are related by an Euler-Rodrigues rotation of an 
%angle $\theta$ in the direction of $\hat {\tilde n} = -\hat i \sin\varphi + \hat j\cos\varphi$ 
%as depicted in Fig. \ref{fig1}, $\bm{\tilde v} = R_{\hat n^\prime}(\theta) \bm v $. 
%We would like to stress the difference between the magnetic axis ($\hat x, \hat y, \hat z$, marked blue in Fig. \ref{fig1}) 
%and the crystallographic ones ($\hat i\parallel$ [110], 
%$\hat j\parallel$ [1$\bar 1$0], $\hat k\parallel$ [001], black arrows of Fig. \ref{fig1}).

Scattering relaxation occurs %within the 2-DEG defined by the crystallographic axis: 
as the $\vec{k}$ vector changes randomly  direction while the modulus remain constant, $|\vec{k}|\simeq k_F$,
with $k_F$ being the Fermi wavevector. \
%During an ESR experiment, the magnetic field direction is changed relative to the
%crystallographic one, such that $\theta, \phi$ determine the azimuthal and polar angles, respectively. 
Classically and neglecting scattering events, the circular motion of the electron is given by
the cyclotron frequency, $\omega_c (\theta)= e B\cos{\theta}/ m$, setting 
$\bm{\tilde k} \simeq k_F(\cos{\omega_c t}\hat i + \sin{\omega_c t} \hat j )$.
In addition,  %relaxation in the 2-DES  via scattering in the plane spanned
%by the crystallographic axis ($\hat i, \hat j$): 
elastic scattering events change randomly the momentum direction in a mean interval $\tau_k$. % of  $\vec{k}$. %with $|\vec{k}|=k_F$, $k_F$ being the Fermi wavevector. \
%This results in a local fluctuating effective magnetic field
%resulting in the  D'yakonov-Perel' (DP) mechanism of spin relaxation \cite{dp}. 
%We describe the scattering processes by the phenomenological parameter $\tau_k$, the mean 
%time electrons move without scattering events. 
%For the typical experiments considered here, we anticipate that the effective spin-orbit 
%In typical experiments, the spin-orbit effective field fluctuates rapidly in the electronic frame due % in the cyclotron or even Larmor time scales due 
%to the motion of the electron and its elastic collisions. % , $\tau_k^{-1} > \omega_l, \omega_c$.  %occurring at a mean time $\tau_k$. 
Due to this cyclotron motion and random scattering, 
the electron `feels' an average  SOC fluctuating field which typically is smaller than its 
instantaneous local value, commonly termed as `motional narrowing' \cite{abragam,schli} 
within the  D'yakonov-Perel' (DP) limit \cite{dp,li_prl}. 
Hence, it appears convenient to express the SOC in terms (\ref{eq_h}) of an effective fluctuating magnetic field  %\cite{li_prl}
in the {\it magnetic} coordinate system 
by a transformation defined in Appendix \ref{app}, namely, an Euler-Rodrigues rotation of all
the (pseudo-) vectors in Eq. (\ref{eq_h}),  $\bm {\tilde\sigma }= R_{\hat {\tilde n}}^\theta \bm \sigma $; 
$\bm{\tilde k} = R_{\hat {\tilde n}}^\theta\bm k$, 
$R_{\hat {\tilde n}}^\theta$ being the rotation matrix of an angle $\theta$ along $\hat{\tilde n}$, as depicted in Fig.\ref{fig1}.
In this rotated frame the time dependency factorizes, %thus in terms of the {\it magnetic} coordinate system, 
\begin{equation}
H^\prime = H_{\rm l} + H_{\rm c}  = \vec \sigma \vec B^{\rm eff}, \quad \vec B^{\rm eff } =  {\rm Re} \{ e^{i\omega_at}\vec Z_a \}, 
\label{eq:Beff}
\end{equation}
\begin{table*}[!hbt]
\caption{\textit{ Components of the effective Rashba ($a =$ R) and Dresselhaus ($a =$  D, 3D) effective magnetic
field along the magnetic axis, $q= x, y, z$.
}}
\label{table1}\centering
\begin{tabular}{|c||c|c|}
\hline
$Z_{a}^q$& $a$ = R (3R), prefactor $\alpha k_F$ ($\alpha_3( k_F^3)$)  & $a$ = D (3D), prefactor $\beta k_F$ ($\beta_3(k_F^3)$)  \\
%\hline
\hline\hline
$q=x$
& $(1-\cos{\theta}) \sin{\varphi}\cos{\varphi} -i[\cos{\theta}\cos^2{\varphi}+\sin^2{\varphi}]$ &
$1-\cos^2{\varphi}(1-\cos{\theta})-i (1-\cos{\theta})\sin{\varphi}\cos{\varphi}$ \\
\hline
%$\cos{\theta}(1-\cos^2{\varphi})+\cos^2{\varphi}-i(1-\cos{\theta} \sin{\varphi}\cos{\varphi} $  \\
$q=y$ &
$\cos{\theta}\sin^2{\varphi}+\cos^2{\varphi}+i(1-\cos{\theta}) \sin{\varphi}\cos{\varphi} $  &
$(1-\cos{\theta})\sin{\varphi}\cos{\varphi} +i[1-\sin^2{\varphi}(1-\cos{\theta})]$ \\ \hline
$q=z$ & $\sin{\theta}( \sin{\varphi} +i\cos{\varphi})$ & $-\sin{\theta}( \cos{\varphi} -i\sin{\varphi})$\\
\hline
%\label{tableB}
\end{tabular}
\end{table*}
with $a = $ R, D, 3R, 3D labeling Rashba, Dresselhaus or cubic SOC components,  
$\omega_R = - \omega_D = \omega_c\cos{\theta}$, and $\omega_{3R} = -\omega_{3D} = 3\omega_c\cos\theta$ (see insets of Fig. {\ref{fig2}}). 
The components of the bi-dimensional pseudo-vector $Z_a$ %in terms of the quantization axis given by $\bm B$, 
along the magnetic axis are obtained in App. \ref{app} and summarized in table {\ref{table1}}. 
We note that in the presence of a magnetic field, the canonical momentum should be replaced by the kinetic one, $\bm k\to\bm k-e\bm{A}$, with 
$\bm{A}$ being the vector potential. However, the additional term is neglectible in typical laboratory frames in high mobility samples. Moreover,
terms in $\bm A$ would  not contribute to the anisotropy in $\varphi$. 
%resulting form an Euler-Rodrigues rotation of all 
%the (pseudo-) vectors in Eq. (\ref{eq_h}), $\bm {\tilde\sigma }= R_{\hat {\tilde n}}^\theta \bm \sigma $; 
%$\bm{\tilde k} = R_{\hat {\tilde n}}^\theta\bm k$, 
%$R_{\hat {\tilde n}}^\theta$ being the rotation matrix of an angle $\theta$ along $\hat{\tilde n}$, as depicted in Fig.\ref{fig1}.

%angle $\theta$ in the direction of $\hat {\tilde n} = -\hat i \sin\varphi + \hat j\cos\varphi$
%The bi-dimensional pseudo-vector $Z_a$, commonly expressed in terms of the crystallographic axis, 
%may be described in terms of its magnetic axis 
%after a rotation by $\theta$, as defined in Fig. {\ref{fig1}}. The components are summarized in table {\ref{table1}}. 
%As the magnetic field is tilted, $|\theta|>0$,  an in-plane magnetic field appears, contributing to both spin relaxation 
% and decoherence (commonly referred to as $T_1$ and $T_2$) processes. 
%On the other hand, we expect that the cyclotron frequency is reduced, enhancing the motional narrowing. 
%\begin{figure}[!hbt]
%\centering\includegraphics[angle=0, width = .45\textwidth]{./D3D}
%\caption{\it \footnotesize
%Spin orientation of the eigenvectors in the presence of
%(a) linear Rashba, (b) linear Dresselhaus, and (c) cubic Dresselhaus.
%When $\vec k$ rotates with a frequency $\omega_c$, the spin orientation rotates
%with (a) $\omega_c$, (b) $-\omega_c$, and (c) $-3\omega_c$.
%}
%\label{figspin}
%\end{figure}
%{\it Method.} 

\subsection{Equations of motion}
\label{sec:methodsB}
In order to evaluate spin relaxation and decoherence, %we utilize the fact that the SOC is weak as compared to $H_0$ and 
we solve the master equations for the density matrix in the interaction representation, 
\begin{equation}
\rho^* (t) = \frac{1}{2}[1 + \vec n(t) \vec\sigma^*(t)], \quad 
\label{eq:defrho}
\end{equation} 
with 
\begin{equation}
o^*(t)= e^{iH_0t}o(t)e^{-iH_0t}, \quad o = \rho, H^\prime, \sigma \label{rhoeq}. 
%H^{\prime^*}(t)&=& e^{iH_0t}H^{\prime *}(t)e^{-iH_0t}, \label{hsoceq} \\
%\sigma^*(t)&=& e^{iH_0t}\sigma(t)e^{-iH_0t} . \label{sigmaeq},
\end{equation}
%where $\rho^*$ is a 2 $\times$ 2 matrix. 
We stress that $\rho^*$ is represented in the magnetic coordinate system. 
This choice is crucial, as it determines the correct quantization axis, 
which  coincides with the direction of $\bm B$ as long the Zeeman energy 
is large compared with the spin-orbit coupling, as occurs in typical ESR 
experiments. %, $g\mu_BB\gg k_F\alpha, k_F\beta$. 
We now derivate (\ref{rhoeq}) to get: 
\begin{align} 
\frac{\ud [n_i(t)\sigma^*(t)] }{\ud t}& = 
\frac{\ud n_i(t)}{\ud t}\sigma^*(t) + 
\frac{i}{\hbar} n_i(t) H_0 e^{iH_0t/\hbar} \sigma_i(t) e^{-iH_0t/\hbar}-
\nonumber \\ &
 - \frac{i}{\hbar} n_i(t) e^{iH_0t/\hbar} \sigma_i(t) H_0 e^{-iH_0t/\hbar}, 
\end{align}
giving:
\begin{equation}
\frac{\ud n_i(t)}{\ud t}\sigma^*(t) = -\frac{i}{\hbar}n_i(t)[H_0,\sigma_i^*(t)] + \frac{\ud [n_i(t)\sigma^*(t)] }{\ud t}
\label{eq:giving}
\end{equation}
On the other hand, we have,  to second order: 
\[
\frac{\ud \rho^*}{\ud t}  \simeq 
\frac{i}{\hbar}[\rho^*_0, H^*_{\rm{SOC}}(t)]
+ \left(\frac{i}{\hbar}\right)^2\int_0^t\left[ 
[\rho^*_0,H^*_{\rm{SOC}}(t^\prime)],H_{\rm{SOC}}^*(t)
\right]\ud t^\prime , 
\]
with $\rho^*_0 = \rho^*(t=0)$. 
Using (\ref{eq:defrho}) and (\ref{eq:giving}), we obtain:
\begin{align}
&\frac{\ud [n_i(t)\sigma_i^*(t)] }{\ud t} \simeq 
\frac{i}{\hbar} n_i(0)[\sigma_i^*(0), \sigma^*_{q}(t)]\overline{B_q^{\rm{eff}}}+ 
\nonumber \\ 
&\quad + \left(\frac{i}{\hbar}\right)^2\int_0^t n_i(0) \left[ 
[\sigma_i^*(0),\sigma^*_{q}(t^\prime)],\sigma_{q^\prime}^*(t)
\right]  {\cal B}_{qq'}(t,t^\prime) 
%\overline{B_q^{\rm{eff}}(t^\prime)B_{q^\prime}^{\rm{eff}} (t) } 
\ud t^\prime. 
\label{eq:me}
\end{align}
with the auto-correlation function, ${\cal B}_{qq'}(t,t^\prime)$ defined as: %and its Fourier transform defined as
\begin{equation}
{\cal B}_{qq'}(t,t^\prime) \equiv
%\int_{-\infty}^{+\infty}\ud\tau
\overline{B^{\mathrm{eff}}_q(t)B^{\mathrm{eff}}_{q'}(t^\prime)}. 
%= \overline{B^{\mathrm{eff}}_q(\tau)B^{\mathrm{eff}}_{q'}(0)}, \ \tau = t-t^\prime. 
 % e^{-i\omega\tau}, 
%\quad b_{q q^\prime}(\omega) = \int_0^\infty {\cal B}_{q q^\prime}(\tau) e^{-i\omega\tau}\ud \tau. 
\label{eq:autoc}
\end{equation} 
We consider in the integral of the right hand of Eq. (\ref{eq:me}) 
ensembles of systems with equal $H_0$, $\rho_0$, but with the effective 
SOC field varying from sample  to sample due to different scattering configurations. 
Hence, for our stationary perturbation $H^\prime$, this ensemble average depends only on 
the interval $\tau = t-t^\prime$. %, and hence the linear term vanishes in Eq (\ref{eq:me}). % vanished. 

On what follows,  we make the following  assumptions:  
%NOte: the Born approximation consist of assuming weak coupling, so decoherence is negligible affected by the interaction, 
%the simplification \rho(0)->\rho(t) means that the evolution equation of the system at t depends only on the present state, 
%and this is the REDFIELD equation, which is local in time and depends on the initial choice of \rho, hence is NOT Markovian. 
%The Markovian approximation comes when we replace t^\prime by t-\tau and let the upper limit of the integral go to infinity.
%It is justified when the time scale associated with the reservoir correlations is much smaller than the time scale over which 
%the state varies appreciably. The Markovian evolution is defined on a coarse-grained time scale, where the dynamical behavior 
%over times of the order of \tau_K are not resolved. Since $\tau_k$ depends on temperature, the Markovian approx. is easily
%justified for weak coupling and high temperature. 
(i) we may use the Redfield approximation \cite{redfield}, replacing $\rho(0) [\sigma(0)]$ 
by $\rho(t)$ $[\sigma(t)]$ in the integrals, such that the evolution equation depends only on the present state, 
%we may replace $H^\prime(t)$ by $H^\prime(0)$ on the right hand side of Eq. (\ref{eqme}), 
(ii) we may use the Markovian approximation, extending the upper limit of the integral to infinity, 
(iii) we may neglect the correlation between $H^\prime$ and $\rho^*(0)$. 
(i) - (iii) are justified as long as 
$\overline{|H^\prime|^2}\sim (H^\prime(0))^2 e^{-t/\tau_k} \ll \hbar^2 \tau_k^{-1} t^{-1}$, 
for times $t$ beyond the transient limit, $t\gg \tau_k$. 
Here, we assumed an exponentially decreasing auto-correlation function 
for the fluctuating spin-orbit field, with the time constant given by the scattering time $\tau_k$. 
Combining Eqs. (\ref{eq:giving}-\ref{eq:autoc}) and the expression in (\ref{eq:H0}): 
\begin{align}
&\frac{\ud n_i(t)}{\ud t}\sigma_i^*(t) = -\frac{i}{\hbar}n_i(t)B_0[\sigma^*_z,\sigma_i^*] - 
\nonumber \\
&-\frac{1}{\hbar^2}\int_0^\infty  n_i(t) \left[ 
[\sigma_i^*(t),\sigma^*_{q}(t-\tau)],\sigma_{q^\prime}^*(t)
\right] {\cal B}_{qq'}(\tau)  \ud\tau. 
\label{eq:ni}
\end{align}
Multiplying Eq. (\ref{eq:ni}) by $\sigma_r^*$, taking the trace, and using the identities:
\begin{equation}
{\rm Tr}\{\sigma_i^*\sigma_j^* \}= 2\delta_{ij} , \quad 
[\sigma^*_i, \sigma^*_j ] = 2i\varepsilon_{ijk}\sigma^*_k,
\label{eq:ids}
\nonumber
\end{equation}
with $\varepsilon_{ijk}$ being the Levi-Civita symbol, we obtain: 
\begin{align}
&\frac{\ud n_i(t)}{\ud t} 2\delta_{ir}
= -  \frac{2i^2B_0}{\hbar} \epsilon_{zip}(2\delta_{pr}) n_i(t) - 
\nonumber \\ &-
\left(\frac{i}{\hbar}\right)^2{\rm Tr}\left\{ 
\int_0^\infty  
\left[[\sigma_i^*(t),\sigma^*_{q}(t-\tau)],\sigma_{q^\prime}^*(t)\right]\sigma_r^*(t)
{\cal B}_{qq'}(\tau) \ud \tau\right\},
%\overline{B_q^{\rm{eff}}(t-\tau)B_{q^\prime}^{\rm{eff}} (t) } \ud \tau\right\}. 
\label{eq:ni2}
\end{align}
The integral of Eq. (\ref{eq:ni2}) is evaluated in appendix \ref{app2}, resulting in 
\begin{equation}
%\frac{\ud n_r(t)}{\ud t} =% \frac{2B_0}{\hbar} \epsilon_{zir} n_i(t) +
\frac{2}{\hbar^2} n_i(t) (\delta_{q^\prime i}\delta_{rk}-\delta_{q^\prime k}\delta_{i r} )
\langle \alpha | \sigma_q |\beta \rangle \langle \beta | \sigma_k |\alpha \rangle 
%e^{-i\omega_{\alpha\beta}\tau} 
b_{qq^\prime}(\omega_{l}),  
%\overline{B_q^{\rm{eff}}(t^\prime)B_{q^\prime}^{\rm{eff}} (t) } \ud t^\prime. 
\label{eqRed}
\end{equation}
with $b_{qq^\prime}$ being the Fourier transform of the autocorrelation function, 
\begin{equation}
b_{q q^\prime}(\omega) = \int_0^\infty {\cal B}_{q q^\prime}(\tau) e^{-i\omega\tau}\ud \tau. 
\label{eq:autoc2}
\end{equation}
%where we have defined: $\omega_{km} = (E_k-E_m)/\hbar$, with $E_k$, $E_m$ being the
%eigenvalues of $H_0$. \
%For our particular case, the eigenvalues are determined by the Larmor frequency, $\hbar\omega_l$ =
%$g\mu_BB_0$, with $g\simeq 2$ \cite{shankar} and $\mu_B = 9.274\times 10^{-24}$J$\cdot$T$^{-1}$,
%the Bohr magneton.
%$\overline{|H^\prime|^2}\sim (H^\prime(0))^2 e^{-t/\tau_k}$. 
%The inequality is easily satisfied by assuming an exponentially decreasing auto-correlation function 
%for the fluctuating spin-orbit field, with the time constant given by the scattering time $\tau_k$, 
%$\overline{|H^\prime|^2}\sim (H^\prime(0))^2 e^{-t/\tau_k}$. 
%Introducing our Hamiltonian in Eq. (\ref{eq:me}), 
%we arrive to a linear evolution of the elements of the density matrix, 
%%\begin{widetext}
%\begin{eqnarray}
%\frac{\ud n_r}{\ud t} &&\simeq
 %-  \frac{2i^2B}{\hbar} \epsilon_{zir} n_i  
%\nonumber \\ 
%%\left(\frac{i}{\hbar}\right)^2
%-\frac{1}{\hbar^2}&&
%{\rm Tr}\left\{ 
%\int_0^\infty  
%\left[[\sigma_i^*(t),\sigma^*_{q}(t^\prime)],\sigma_{q^\prime}^*(t)\right]\sigma_r^*(t)
%{\cal B}_{qq'}(t^\prime)
%%\overline{B_q^{\rm{eff}}(t-\tau)B_{q^\prime}^{\rm{eff}} (t) } 
%\ud t^\prime\right\}, 
%\nonumber \\
%\label{eq:ni2}
%\end{eqnarray}
%\end{widetext}
Eq. (\ref{eq:ni2}) describes then a linear evolution of the elements of the density matrix, 
\begin{equation}\label{eq:Bloch}
%\frac{\ud \vec{n}(t)}{\ud t} 
\dot{\vec{n}}
= %\frac{4}{\hbar^2} 
-A \vec{n}
\end{equation}
with 
\[ A = 
\left(
\begin{array}{ccc} 
b_{yy}(\omega_l) +b_{zz}(0) %+ib_{xy}(\omega_l)-ib_{yx}(\omega_l) 
&  B_0\hbar- b_{xy}(\omega_l) & -b_{xz}(\omega_l) \\ 
B_0\hbar -b_{xy}(\omega_l) & b_{xx}(\omega_l) +b_{zz}(0)  & -b_{yz}(\omega_l)\\
-b_{zx}(0) & -b_{zy}(0) & b_{xx}(\omega_l) +b_{yy}(\omega_l).
\end{array}
\right)
\]
%where Larmor frequency is $\omega_l$ = $g\mu_BB_0/\hbar$, with $\mu_B$ being the 
%Bohr magneton. %, with $g\simeq 2$ \cite{shankar}. 
%The relaxation times are $T_i^{-1} = -{\rm Re}\{\lambda_i\}$, where $\lambda_i$ are the eigenvalues
%of $A$. 
In general, the eigenvalues of $A$ will consist of a complex conjugate pair and a real number.
$T_1^{-1}$ corresponds to the real one and $T_2^{-1}$ to the real part of the conjugate pair.
In this sense, we may disregard the first term on the right of Eq. ({\ref{eq:ni2}}), 
which would only contribute to the imaginary part of the eigenvalues of (\ref{eq:Bloch}). 

Eq. (\ref{eq:Bloch}) corresponds to a general result within  Redfield's approach \cite{abragam, schli} 
that describes the time evolution of 
the spin components, which ultimately yields dephasing and relaxation spin times. 
In the next sections, we employ this approach to derive $T_1^{-1}$ and $T_2^{-1}$ 
in experimentally relevant systems. 

\subsection{Decoherence and relaxation in typical ESR experiments}
\label{sec:methodsC}
We evaluate next the autocorrelation functions ${\cal B}_{qq'}(\tau)$ and its Fourier transform, $b_{qq^\prime}$ 
in typical samples of our interest, with effective fields given by (\ref{eq:Beff}).
On these systems, we can assume that ${\cal B}_{qq'}(\tau)$ is an even function of time, such that 
$b_{qq^\prime}(\omega) = b_{qq^\prime}(-\omega)$, and that the 
$x$-, $y$- and $z$-components of the field fluctuate independently. 
Noting that the time dependency of $B_q^{\rm eff}$ is determined by the scattering processes and the 
cyclotron motion, we can write, for the linear contributions: 
\begin{widetext}
\[
{\cal B}_{qq'}(\tau) \simeq \overline{ (Z_{q,D}e^{i\omega_{c}\tau/2} + Z_{q,R}e^{-i\omega_{c}\tau/2})
(Z_{q^\prime,D}^*e^{i\omega_{c}\tau/2} + Z_{q^\prime,R}^*e^{-i\omega_{c}\tau/2})
} e^{-|\tau|/\tau_k}, 
\]
\end{widetext}
leading to:  
\begin{align}
{\cal B}_{qq'}(\tau) &\simeq& \left[ 
{ Z_{q,D} Z_{q^\prime,D}^*} e^{i\omega_{c}\tau} +
{ Z_{q,D} Z_{q^\prime,R}^*} +
{ Z_{q,R} Z_{q^\prime,D}^*}+ 
\right.\nonumber \\ &&+\left. 
{ Z_{q,R} Z_{q^\prime,R}^*} e^{-i\omega_{c}\tau} 
\right] e^{-|\tau|/\tau_k}, 
\end{align}
and the Fourier transform, %for  ${\cal B}_{qq'}(\tau)$ even in time and independently-fluctuating components, gives:
%of the autocorrelation functions: 
\begin{align}
b_{qq^\prime}&\simeq&
\left(\frac {{ Z_{q,D} Z_{q^\prime,D}^*}\tau_k }{1 + (\Omega_-\tau_k)^2}
+\frac {{ Z_{q,R} Z_{q^\prime,R}^*}\tau_k }{1 + (\Omega_+\tau_k)^2}+
\right.\nonumber \\ &&\left.
+\frac {[{ Z_{q,D} Z_{q^\prime,R}^*} +{ Z_{q,R} Z_{q^\prime,D}^*}]\tau_k }
{1 + (\omega_l\tau_k)^2}\right)\delta_{qq^\prime}, 
\end{align}
with $\Omega_\pm = \omega_l\pm\omega_c$.  
%Let us focus first on the linear-in-$k$ contributions, $H^\prime$.
We stress that the different terms for the auto-correlation function are not present in other approaches \cite{glazov}, 
and as we will see, may boost the anisotropy on the decoherence in some cases.

\subsubsection{Perpendicular magnetic fields}
We first focus on the most straight-forward case,
where the external magnetic field is parallel to the growth direction $\hat k$,  $\theta =0$. 
The crystallographic and magnetic coordinate systems coincide and hence, 
fluctuations in the effective magnetic field 
occur only along $\hat x$- and $\hat y$-directions ($b_{zz}(\omega) =0$ ), preserving the U(1) symmetry: 
$Z_{x,R}(\theta=0) = -i\alpha k_F $ ,
$Z_{x,D}(\theta=0) = \alpha k_F $,
$Z_{y,R}(\theta=0) = \beta k_F $ and $Z_{y,D}(\theta=0) = i\beta k_F $, with the
remaining $Z_{q,R/D}$ being 0, bearing $b_{xy}(\omega) = b_{xz}(\omega) = b_{zz}(\omega) =0$. 
We obtain: 
\[
b_{xx}(\omega_l)  = b_{yy}(\omega_l) = k_F^2\tau_k\left(\frac{\beta^2}{1+(\Omega_-\tau_k)^2} + \frac{\alpha^2}{1+(\Omega_+\tau_k)^2}\right). 
\]
Eq. (\ref{eq:Bloch}) contains only diagonal terms, giving an straight forward expression for $T_{1,2}$: 
%(note that magnetic $\hat x$-, $\hat y$- $\hat z$-axis coincide with the 
%crystallographic ones, $\hat i$, $\hat j$ and $\hat {\underline k}$, respectively). 
%Neglecting the cubic terms in the Dresselhaus SOC, we have, 
\begin{equation}
T_1^{-1}(\theta=0) =  \frac{8\tau_k k_F^2}{\hbar^2}  \left(\frac{\beta^2}{1+(\Omega_-^0\tau_k)^2} + \frac{\alpha^2}{1+(\Omega_+^0\tau_k)^2}\right), 
%\frac{4\tau_k(\beta k_F)^2(1+\gamma^2)}{\hbar^2(1+(\omega_{-}\tau_k)^2)}, 
%\quad{\rm and}\quad
%T_2^{-1}(0) =\frac{T_1^{-1}(0)}{2} %\frac{4\tau_k(\beta k_F)^2(1+\gamma^2)}{\hbar^2(1+(\Omega_{-}\tau_k)^2)}. 
%=\frac{ T_1^{-1}(0)}{2}. 
\label{eq:ttwo0}
\end{equation}
with %$\beta = \gamma \langle k_z^2 \rangle$ and 
$\Omega_\pm^0 = \omega_c^0 \pm \omega_l$, $\omega_c^0\equiv \omega_c(\theta=0)$. % and $T_2^{-1}(\theta=0) = T_1^{-1}(\theta=0)/2$. 
%Viewing the problem from the rotating frame of the effective Rashba (Dresselhaus) magnetic field
%at $\Omega_- =\omega_l - \omega_{c}$ ($\Omega_+ = \omega_l +\omega_c)$, these 
%results are reasonable, since the $T_1$ corresponds to a change in the $z$-magnetization.
%Such change is brought about by `static' fields in either the $x$- or $y$-directions in the rotating frame.
%But `static' $x$- or $y$-fields in the rotating frame oscillate at $\Omega_-$  ($\Omega_+$)in the laboratory frame.
This result  reflects the additive contributions of spin relaxation  due to Rashba and Dresselhaus.  %$T_1$ corresponds to a change in the $z$-magnetization.
%Similar results had already been reported for $\theta =0$ \cite{glazov}. 
%Such change is brought about by fluctuating fields in either the $x$- or $y$-directions. 
The Rashba (Dresselhaus) effective fields in the electronic frame fluctuate at 
$\Omega_+^\theta =(\omega_l + \omega^0_{c}\cos{\theta})$ ($\Omega_-^\theta = |\omega_l -\omega^0_c\cos{\theta})|$. 
As a result,
 $T_{1,2}^{-1}$ has a minimum as a function of $\tau_k$ when $\Omega_-\tau_k = 1$.
Viewing the problem from the rotating frame of the effective magnetic field
(the one rotating at $\Omega_- =\omega_l - \omega_{c}$,
which is the resulting rate of change of the effective magnetic field due to SOC), these
results are reasonable, since $T_1$ corresponds to the time it takes to change the $z$-magnetization.
Such change is brought about by `static' fields in either the $x$- or $y$-directions in the rotating frame.
But `static' $x$- or $y$-fields in the rotating frame oscillate at $\Omega_-$ in the laboratory frame.
If the scattering time $\tau_k$ is comparable to $\Omega_-^{-1}$, then the $T_1$ process become most
effective, as the electron has time to `feel' the effects of the change in $x$- or $y$-fields.
%Clearly, for $\omega_c =\omega_l$, the Rashba field appears static, and the equations above would not 
%be valid, so we focus on the experimentally relevant limit, $\omega_c\gg\omega_l$. 
In the rapid motion limit, $\Omega_-\tau_k, \omega_{c}\tau_k \ll 1$, we note that
$T_i^{-1}\propto \tau_k$, i.e., the shorter $\tau_k$ (that is, the more \emph{ rapid} the motion), the
narrower the resonance. This phenomenon is therefore called \emph {motional narrowing}: The motion
narrows the resonance because it allows a given spin to sample many fields, some of which cause it to advance
in phase; others, to be retarded. The dephasing takes place, then, by a random walk of small steps, each one much
less than a radian.
It is  well known that   in this limit, 
$T_2^{-1}(0) \simeq T_1^{-1}(0)/2$, due to the absence of fluctuations of the effective field along $\hat z$. 

In contrast, when there is no `motion', a given spin experiences a constant local field.
Each collision gives a loss in phase memory, and thus a more rapid collision rate
produces a shorter phase memory and a broader line, thus termed as `{\it collision broadening'}.
 $T_2^{-1}$ would then be proportional to the collision rate, $\tau_k^{-1}$, which is in clear contrast with 
the motional narrowing: the phase of the oscillation is then changed by each collision. 
Clearly, for $\omega_{c} = \omega_{l}$, the SOC field appears static,  and the DP model breaks down.  %for $\omega_{c} = -\omega_{l}$.
On typical experimental setups, we have that $\omega_{c}(0) \gg \omega_{l}$, however, as 
$\theta\simeq \pi/2$, the collision broadening limit can be reached, as we will see. 

\subsubsection{Arbitrary magnetic fields}
\begin{figure}[!hbt]
\centering\includegraphics[angle=0, width = .46\textwidth]{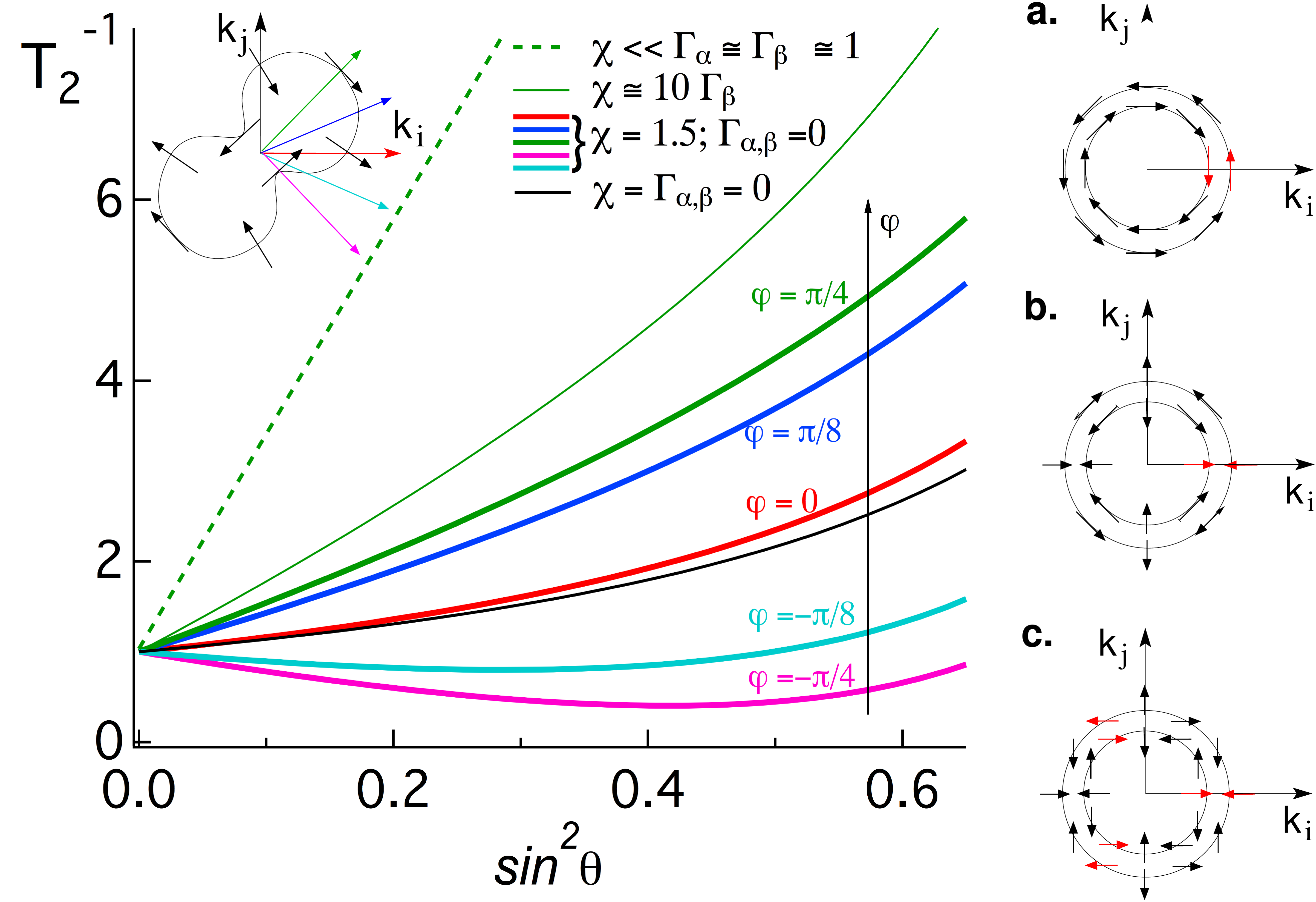}
\caption{\it \footnotesize  $T_2^{-1}/T_2^{-1}(0)$ 
as a function  $\sin^2{\theta}$, $\theta = \measuredangle(\hat k, \hat z)$, for  five 
orientations of  $\bm B$ %$B_\parallel$ relative 
relative to the crystallographic axis,  $\varphi = 0, \pm \pi/8, \pm \pi/4$ in the linear SOC limit
(thick solid lines) % $\chi = 1.5$, $\alpha_3, \beta_3 = 0$. 
, including moderate cubic terms (thin green curve),  % $\alpha, \beta \sim \alpha_3, \beta_3$, $\varphi = \pi/4$.
and large cubic terms (broken green trace)% was obtained including large cubic terms.  %$\alpha, \beta \ll \alpha_3, \beta_3$, $\varphi = \pi/4$.
%We assume a sample characterized by $\tau_k$ = 10 ps.
Left inset: magnitude (solid black curve) and direction (black arrows) of the effective fluctuating 
field in $\bm k$ space. The colored arrows mark the direction of $\bm B_\parallel$.
Right: Spin orientation of  eigenvectors for
(a) linear Rashba, (b) Dresselhaus, and (c) cubic Dresselhaus.
%When $\vec k$ rotates with a frequency $\omega_c \hat k$, the spin orientation rotates
%with (a) $\omega_c \hat k$, (b) $-\omega_c \hat k$, and (c) $-3\omega_c \hat k$. 
}
\label{fig2}
\end{figure}

As the external magnetic field $\bm B$ is tilted with respect to the crystallographic axis,  
the projections of the electronic movement in $\hat i, \hat j$ onto $\hat z$ result in 
effective fluctuating fields along $z$-axis, resulting in anisotropic contributions: 
%Hence, we expect decoherence to be anisotropic. 
%Assuming that the $x$- $y$- and $z$ components of the field fluctuate independently, 
%$b_{ij}(\omega)=0$  = 0 if $i\neq j$, we have: 
\begin{equation}
\label{eq:ttwo}
T_2^{-1} \simeq \frac{2}{\hbar^2} (b_{xx}{(\omega_l)}+b_{yy}{(\omega_l)} +2 b_{zz}(0) ), 
\end{equation}
where we have assumed that the components of the effective field fluctuate independently, 
$b_{\alpha\alpha^\prime} = 0, \alpha\neq\alpha^\prime$.
%with $b_{zz}(0)$ arising from the projections of the fluctuations into the $\hat z$ axis. %These fluctuations are due to the electronic movement in the 2-DEG. 
Using Eqs. (\ref{eq:autoc}), (\ref{eq:autoc2}) and table \ref{table1}, we have: 
\[
b_{zz}(0)  =  %\tau_k\frac{|Z_R^z|^2 + |Z_D^z|^2| + 2{\rm Re}\{ Z_R^z Z_D^{z*}\} }{1+(\omega_c\tau_k)^2} = 
\tau_k\frac{k_F^2\sin^2{\theta}(\alpha^2+\beta^2+\alpha\beta\sin{2\varphi}) }{1+\omega_c^2\tau_k^2\cos^2{\theta}}. 
\]
$b_{zz}(0)$ is maximal along the [$110$] direction ($\varphi = 45^\circ$), 
where both contributions of SOC are parallel, and minimal along 
[$1\bar10$], ($\varphi = -45^\circ$) where these are anti-parallel (see inset of fig. \ref{fig2}). 
Note that the fluctuations occur with a frequency given by $\omega_c$ in the electronic frame.
%Thefluctuate with $\omega_c$ in the frame of the electron. 
%Note that this factor is $\propto [1\pm (\alpha/\beta)^2]$, 
The other two terms can be combined to give: 
\begin{eqnarray}
b_{xx}(\omega_l) + b_{yy}(\omega_l) &=& 2\tau_k\left( 
\frac {{ |Z^x_{D}|^2 + |Z^y_{D}|^2} }{1 + (\Omega_-^\theta\tau_k)^2}
+\frac {{ |Z^x_{R}|^2 + |Z^y_{R}|^2} }{1 + (\Omega_+^\theta\tau_k)^2}
\right.\nonumber \\ &&\left.
+\frac { 2{\rm Re}\{ Z^x_{D} Z^{x*}_{R} +  Z^y_{R} Z^{y*}_{D}\}  }{1 + (\omega_l\tau_k)^2} 
\right). 
\label{eqk}
\end{eqnarray}
The first two  terms on the right appeared already in Eq. (\ref{eq:ttwo0}),
pertaining to the fluctuations of the effective magnetic field along $x$- and $y$- axis. 
Looking at  table \ref{table1}, we find $|Z^x_{D(R)}|^2 + |Z^y_{D(R)}|^2 = \beta^2(\alpha^2)k_F^2 (1+\cos^2{\theta}) $. 

The last term of Eq. (\ref{eqk}) reflects an interference effect, % of both SOC fluctuations, 
arising due to the correlations  of the linear Rashba and Dresselhaus effective field, for which 
the cyclotron precession is canceled in the electronic frame. 
%${\rm Re}\{ Z_{x,D} Z_{x,R}^* +  Z_{y,R} Z_{y,D}^*\} = 2\alpha \beta
%\sin^2{\theta}\cos{2\varphi}$. % appearing at finite $\theta$ and $\varphi$. 
%evolving with $\omega_l$. 
This highly anisotropic term 
(${\rm Re}\{ Z_{x,D} Z_{x,R}^* +  Z_{y,R} Z_{y,D}^*\} = 2\alpha \beta\sin^2{\theta}\cos{2\varphi}$ ) 
results from the fluctuations of the projections of the SOC 
onto the $z$-axis, hence proportional to $\sin^2{\theta}$. 
It has a maximal (minimal) value for $\varphi=\pi/4$ ($\varphi=-\pi/4$). 
Although a term with some similarities on the angular dependency was obtained by Glazov {\it et al.} \cite{glazov}, 
the overall expression differs substantially with our results, and also a 
connection with effective fluctuating fields or motional narrowing was not provided in \cite{glazov}.

The anisotropy can be quantified, in the linear SOC limit, in terms of a single parameter,
$\chi=\beta/\alpha$, 
suggesting a method to determine the relative strength of both linear SIA and BIA couplings using ESR: 
 %and the 
%angle $\varphi$ between  $B_\parallel$ and $\hat i$, 
\begin{widetext}
\begin{equation}
\frac{T_2^{-1}(\theta,\varphi)}{T_2^{-1}(0)} \simeq
\frac{1}{2\nu}\left[ (1 + \cos^2{\theta})\left(
\frac{1}{ 1 + (\Omega_+^\theta\tau_k)^2} 
+\frac{\chi^2}{1 + (\Omega_-^\theta\tau_k)^2} \right)
+ \sin^2{\theta}\left ( \frac{2\chi\sin{2\varphi}}{1 + (\omega_l\tau_k)^2}
+\frac{2(1+\chi^2 + \chi\sin{2\varphi})}{1 + \omega_c^2\tau_k^2\cos^2{\theta}}
\right)
\right]
\label{eq:ttwonorm}
\end{equation}
\end{widetext}
with $\nu = \hbar^2 T_1^{-1}(0)/(\tau_k k_F^2\alpha^2)$ and $T_2^{-1}(0) = T_1^{-1}/2$.  %given in (\ref{eq:ttwo0}). 
%Hence, an estimation of $\chi$ becomes possible in typical ESR experiments. ==
%In the linear-in-$k$ limit, $\chi$ %the relative strength of the Rashba and Dresselhaus SOC couplings 
%can thus be determined by ESR. 

\subsubsection{Decoherence including cubic SOC terms}
So far we did not take into account higher-in-momentum terms for the SOC. 
As it has been recently reported in literature, cubic terms can be  
large compared to the linear terms in narrow QWs \cite{kohda}. 
%If a cubic term, say $\beta_3$ dominated, $\beta_3 \gg \alpha, \beta$ we would formally obtain 
%a result analogous to Eq. (\ref{eq:ttwonorm}), only changing $\omega_c \to 3\omega_c$.
%As a result, the anisotropy may increase when the cubic terms are included.
%This can be easily seen by writing the last term of Eq. (\ref{eq:ttwonorm}),
%The last term on Eq. (\ref{eq:ttwonorm}) would read 
%$a\sin^2{\theta}/(1+9\omega_c^2\tau_k^2\cos^2{\theta})$, with $a$ 
%being a function of $\beta_3, \chi$ and $\varphi$. 
%The factor 9 in the denominator %{\large .....}  
%causes an effective suppression of motional narrowing as $\theta$ increases, 
%resulting in pronounced anisotropies. 
Even if these are small compared with the linear terms, interference effects 
due to simultaneous SOC sources are highly anisotropic.
Moreover, the effective fields precess faster during the cyclotron motion of the electron,   
suggesting that a full derivation of the decoherence 
including all SOC  is necessary. Defining the dimensionless relative strengths, %When different SOC occur simultaneously, interference effects have to be considered. 
%We obtained the expression for the angular dependency of 
%electron spin decoherence, including the cubic SOC terms with relative strengths 
$\Gamma_\alpha = \langle k_i^2\rangle \alpha_3/\alpha$ and $\Gamma_\beta = \langle k_i^2\rangle \beta_3/\alpha$, we get:  
\begin{widetext}
\begin{eqnarray}
\frac{T_2^{-1}(\theta,\varphi)}{T_2^{-1}(0)} \simeq
\frac{1}{2\nu}\left[ (1 + \cos^2{\theta})\left(
\frac{1}{ 1 + (\Omega_+^\theta\tau_k)^2} 
+\frac{\chi^2}{1 + (\Omega_-^\theta\tau_k)^2} 
+\frac{\Gamma_\alpha}{ 1 + (\Omega_{2+}^\theta\tau_k)^2} 
+\frac{2\chi\Gamma_\beta}{ 1 + (\Omega_{2-}^\theta\tau_k)^2} 
+\frac{(\Gamma_\alpha)^2}{ 1 + (\Omega_{3+}^\theta\tau_k)^2} 
+\frac{(\Gamma_\beta)^2}{ 1 + (\Omega_{3-}^\theta\tau_k)^2} 
\right)\nonumber \right.\\ +\left.
\sin^2{\theta}\left ( \frac{2(\chi +\Gamma_\alpha\Gamma_\beta)\sin{2\varphi}}{1 + (\omega_l\tau_k)^2}
+\frac{2 \Gamma_\beta\sin{2\varphi}}{1 + (\Omega_-^\theta\tau_k)^2} 
+\frac{2\chi \Gamma_\alpha\sin{2\varphi}}{1 + (\Omega_+^\theta\tau_k)^2} 
+\frac{2(1+\chi^2 + \chi\sin{2\varphi})}{1 + \omega_c^2\tau_k^2\cos^2{\theta}}
+\frac{2\chi(\Gamma_\alpha + \Gamma_\beta)}{1 + 4\omega_c^2\tau_k^2\cos^2{\theta}}
+\frac{(\Gamma_\alpha)^2 + (\Gamma_\beta)^2}{1 + 9\omega_c^2\tau_k^2\cos^2{\theta}}
\right)
\right],
\nonumber\\
\label{eq:ttwonorm2}
\end{eqnarray}
\end{widetext}
with $\Omega_{n\pm}^\theta = n\omega_c\cos{\theta}\pm \omega_l$, $n=2,3$, and $\nu$ determined now by a relaxation time 
that includes cubic contributions. 
Previous work on the angular dependency of $T_2$ yielded a different expression (see Eqs.  
(10), (11) and (15) of Glazov  {\it et al.} \cite{glazov}), 
where a classical field approach for the kinetic equations within the density matrix formalisms resulted 
in $\chi$ and $\varphi$-independent expressions for $T_2$ without third harmonic contributions. 
Our results yield a more general theory, where the auto-correlations of the effective fluctuating fields 
are explicitly taken into account. 
As we will see below, the experimental data indicates that the cubic contributions as well as interference 
terms resulting from the field correlations are critical for an accurate description of  decoherence rates. 

\section{Results} 
\label{sec:results}
\subsection{Theoretical results}
\label{sec:resultsA}
Figure \ref{fig2} shows the angular dependency of the normalized decoherence, $T_2^{-1}(\theta)/T_2^{-1}(0)$, 
for linear SOC, $\chi = 1.5 \gg \Gamma_{\alpha,\beta}$ (solid colored lines), 
moderate cubic SOC, $\chi =1.5 \sim 2\Gamma_{\alpha,\beta}$ (thin green curve) 
and large cubic SOC, $\Gamma_{\alpha,\beta} \gg \chi$ (broken green curve). 
The black curve, for comparison, corresponds to $\chi =0$, hence spin-rotation 
symmetry is only partially broken down to U(1),  and the angle $\varphi$ is irrelevant: 
The effective magnetic field has the same magnitude along any direction in the 2DEG (see Fig. \ref{fig2}a.). 
The results illustrate that $T_2$ processes are greatly suppressed in the presence of a
perpendicular field, as the electron can circle around with many scattering events,
averaging out the effective magnetic field of SOC. %\ On the other hand,
As the magnetic field is tilted, the cyclotron frequency decreases, suppressing motional narrowing.
Suppression of motional narrowing results in an increase of decoherence, increasing $T_2^{-1}$.
%\[\nu =\frac{1}{1+(\Omega_+(\theta=0)\tau_k)^2} + \frac{\gamma^2}{1+(\Omega_-(\theta=0)\tau_k)^2} \]
%The $\varphi = 0$ case (red curve) is equivalent to a fit with only one source of SOC %($\alpha$-fit), 
%thus marking a lower limit in the SOC-induced anisotropy. 
%When both BIA and SIA are comparable, a  (green curve, for $\varphi = \pi/20$, 
%orange for $\varphi = \pi/8$, and blue, $\varphi = \pi/8$). 
%The colored curves demonstrate the relevancy of the angle $\varphi$:  $T_2^{-1}$ 

In the linear SOC limit with both Rashba and Dresselhaus coupling, 
we observe that the anisotropy is remarkable, with $T_2^{-1}$ 
differing by an order of magnitude at $\theta = \pi/4$ ($\sin^2{\theta}=0.5$) for 
two different crystallographic directions of $B_\parallel$ %, given by $\varphi$ 
(magenta, $\varphi = -\pi/4$ and green, $\varphi = \pi/4$). 
The solid 8-shaped curve in the left inset corresponds to the magnitude of the fluctuating (SOC) 
field in the linear limit (note that it is maximal along $\varphi =\pi/4$ and minimal along $\varphi =-\pi/4$), 
while the black arrows mark the corresponding direction of this effective field in $k$-space. 
The colored arrows represent the $B_\parallel$ direction in the crystallographic plane for the five 
chosen directions, $\varphi =0, \pm\pi/8, \pm\pi/4$. 
As moderate cubic terms are included, the anisotropy is enhanced (thin green curve).
The broken green curve corresponds to the limit
$\Gamma_{\alpha,\beta}\gg\chi\sim 1.5$, when the cubic terms dominate anisotropy. 
%We plot for comparison the $\chi =0$ case, for which all the  directions of $B_\parallel$ are equivalent 
%by symmetry, and hence, no $\varphi$ dependency is observed: the magnitude of the effective 
%SOC field is constant (see Fig. \ref{fig2} a). 
%We note that the $\varphi = 0$ fit (red curve) is almost equivalent to a linear fit with $\chi=0$, {\it i.e.}, with 
%only one linear SOC term, as the terms in $\chi$ cancel out in the normalized expression, 
%$T_2^{-1}(\theta, \varphi=0),  T_2^{-1}(0) \propto (1+\chi^2)$.  

%For $\beta, \alpha \ll \beta_3, \alpha_3$,  decoherence is enhanced due to 
%the relative faster suppression of motional narrowing as $\theta$ is varied. 
%The broken curve corresponds to $\beta \sim \alpha \sim \beta_3 \sim \alpha_3$

\subsection{Relation to experiments} 
\label{sec:resultsB}
We now analyze existing experimental data, \cite{truitt,graeff,wilamowski,tyryshkin}, by taking into account 
the angular anisotropies introduced in Eq. (\ref{eq:ttwonorm2}) and employing the relative couplings $\Gamma_{\alpha,\beta}, \chi$ as 
fitting parameters. 
We focus on experimental results in Si/SiGe heterostructures, since there exist a number of experiments with unexplained anisotropy. 
In these samples, the 2-DEG is formed within a $l$ nm thick strained Si layer  grown on a strain-relaxed Si$_{1-x}$Ge$_{x}$, $x$ = .25-.35.  
We note that the main parameter characterizing the samples is $\tau_k$, obtained from electron mobility data on the 
pertaining references. 
In these experiments, the scattering time shall be  compared with any other time scales for typical experimental fields (~1 T). % when the field is parallel to the growth direction,
%the scattering time (this varies from sample to sample) is of the order of a few ps,
The cyclotron time is ($\tau_c =2\pi/\omega_c$) $\sim$ 10ps (note that this one increases with $\theta$),
and the "Larmor time" is ($\tau_l =2\pi/\omega_l$) $\sim$ 10$^2$ ps. \
%Eqs. (\ref{eq:ttwonorm}-\ref{eq:ttwonorm2}) apply as long as $\tau_k\lesssim\tau_c,\tau_l$ 
Further details on each sample are given on table \ref{tab}. 
\begin{table}[!hbt]
\caption{\textit{ 
Transport parameters for the experimental data considered in this work: 
momentum scattering time, $\tau_k$, 
average value of $g$-factor, 
width of Si QW, $l$, and carrier concentration $n_e$. 
}}
\label{tab}\centering
\begin{tabular}{|c||c|c|c|c|}
\hline
Sample & $\tau_k$ [ps] & $g$-factor & Si QW $l$ [nm]  & $n_e$ [$\times$10$^{11}$cm$^{-2}$]\\
\hline
Ref. \cite{graeff} & 10.2 &2.0005 & 15 &1.0
\\ \hline
Ref. \cite{wilamowski} & 10&2.0005 & 20 & 3.0
\\ \hline
Ref. \cite{tyryshkin_prl} & 10.2 & -- & 20  & 3.0
\\ \hline
UW-30903& 9.4&2.0005 & 10&4.3
\\ \hline
UW-030827 & 9.7 & 2.0013 &10 &4.8
\\ \hline
IBM-01 & 4.3 & 2.00013  & 8&4.0
\\ \hline
UW-31203& 1.8 &2.0003 & 10&2.6
\\ \hline 
UW-31124& 6.9&2.0012 & 10&4.7
\\ \hline
UW-31121& 5.0&2.0013 & 10&5.4
%\\ \hline
%Ref. [\cite{jantsch_PE02}]& & & 
\\
\hline\hline
\end{tabular}
\end{table}

\begin{figure}[!hbt]
\centering\includegraphics[angle=0, width = .4\textwidth]{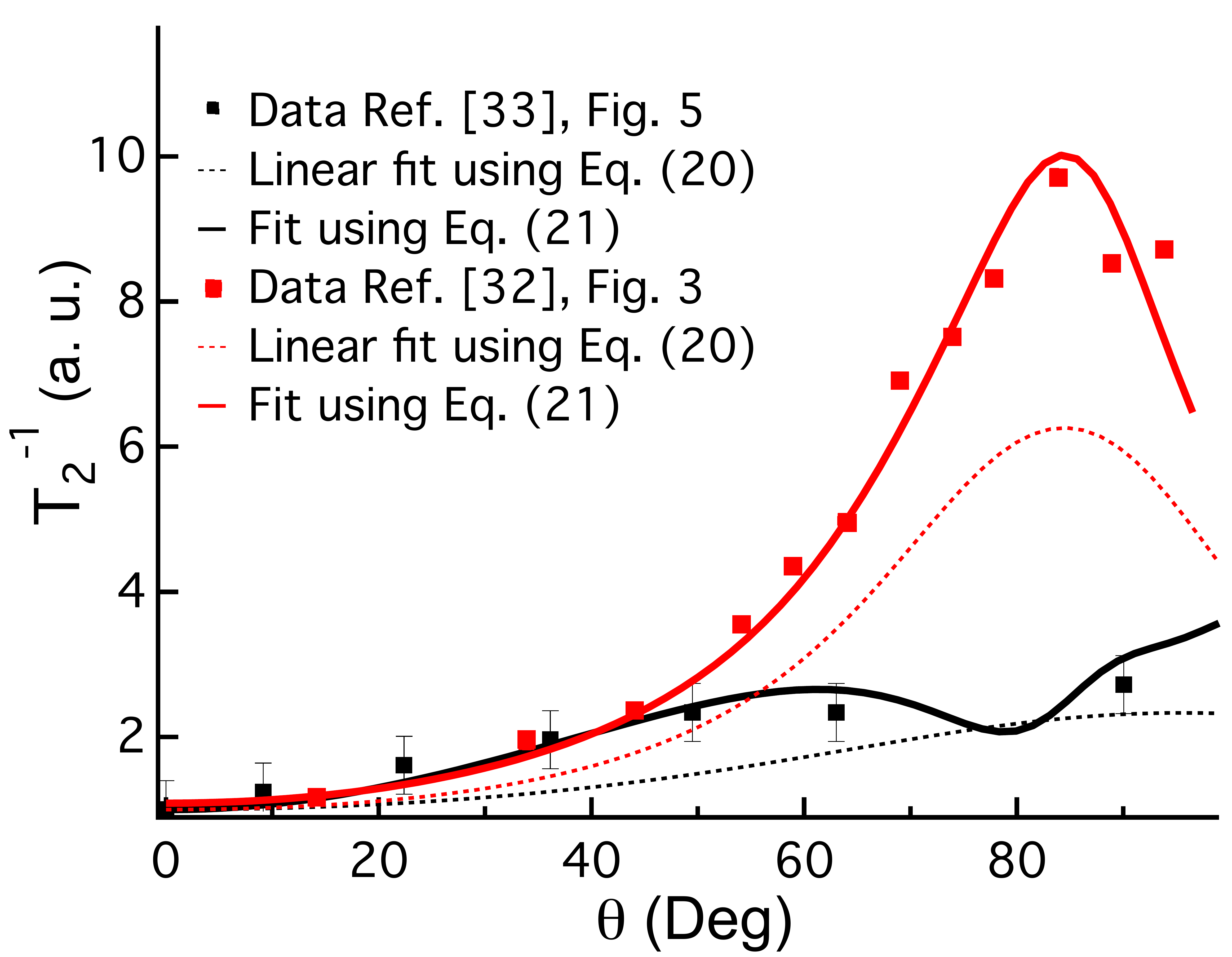}
\caption{\it \footnotesize Experimental data (squares), theoretical fits including cubic terms (solid curves), and fits including linear SOC terms (broken curves). 
Black: Ref. \cite{graeff}, fitted with $\chi = 0.8$ and $\Gamma_\beta = 0.2$ 
Red: Ref. \cite{wilamowski}, fitted with $\chi = 10$ and $\Gamma_\beta=2$. 
}
\label{figres2}
\end{figure}

Fig. \ref{figres2} shows extracted data from Refs. \cite{graeff} (black) and \cite{wilamowski} (red), respectively. 
The solid lines are theoretical fits using Eq. (\ref{eq:ttwonorm2}), whereas the broken ones are fits including only linear SOC terms, as in Eq. (\ref{eq:ttwonorm}). 
For the former case, a small cubic term $\Gamma_\beta = 0.2$ allows us to reproduce the experimental data of Graeff {\it et al.}, with $\beta = .8\alpha$. 
%the anisotropy induced by linear terms irrelevant.  
The black lines, however, fit the data of Wilamowski {\it et al.} with  $\chi = 10$ and $\Gamma_\beta=2$, coinciding with the 
parameters used by Tyryshkin {\it et al.} \cite{tyryshkin_prl} (not shown). This would imply $\Gamma_\beta = .2\beta$ and $\beta = 10\alpha$, 
that is, a system dominated by linear Rashba and with a sizable cubic contribution. 
One data point from Jantsch {\it et al.} would be consistent with $\chi=1.5$ and a large cubic term, $\Gamma_\beta=7.5$. 
We stress that we do not attempt to extract the relative coupling strengths, $\chi, \Gamma_{\alpha,\beta}$, 
as different combinations of the three parameters could be consistent with the data. % (also the angle $\varphi$). 
A more reliable determination of the parameters would require  $\varphi$-resolved measurements. 
However, absence of cubic terms in the SOC would render impossible a fit to the data analyzed so far, for DP-relaxation.   

\begin{figure*}[!hbt]
\centering\includegraphics[angle=0, width = .7\textwidth]{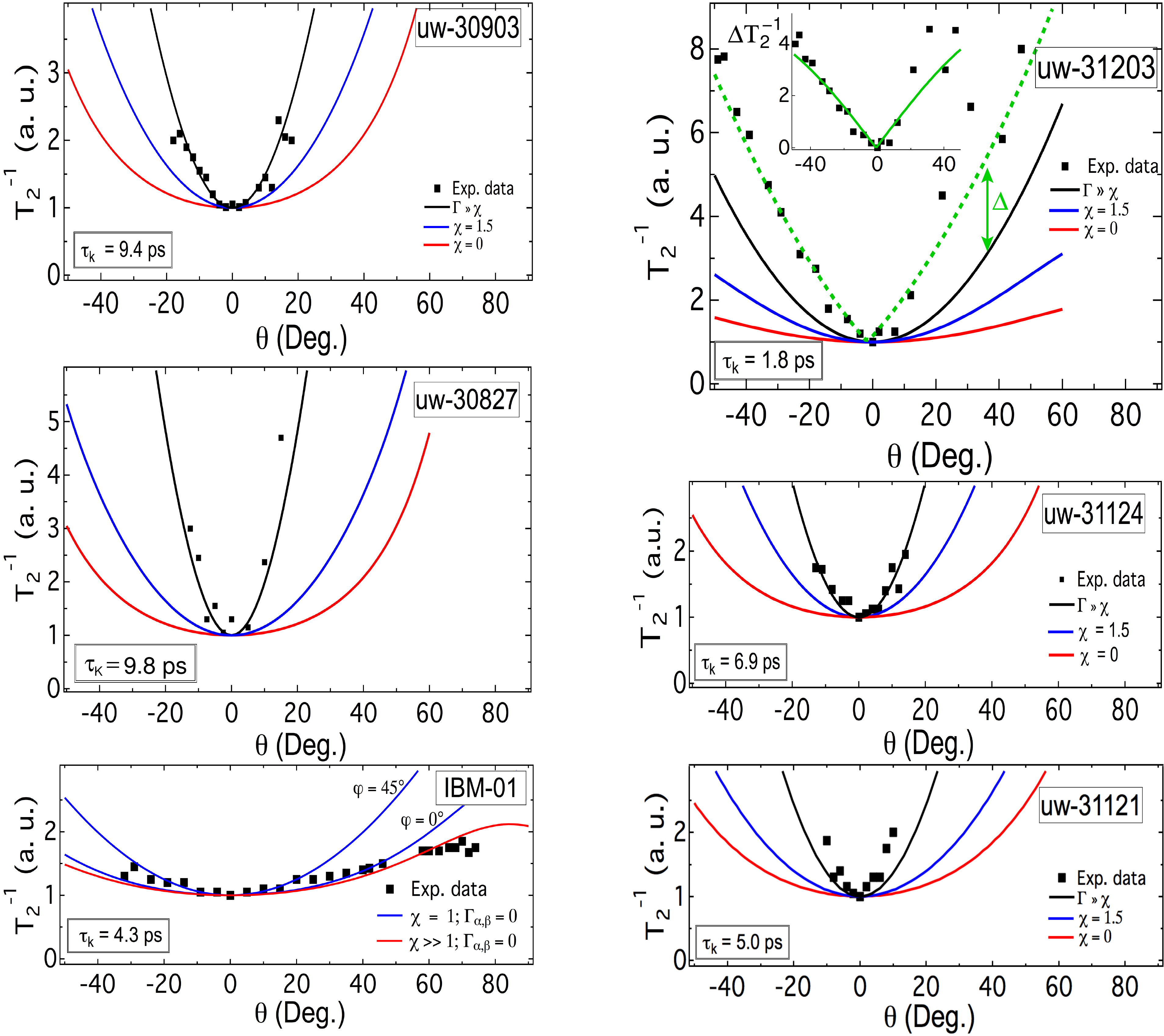}
\caption{\it \footnotesize Experimental data (squares) and theoretical fits (solid curves), including
Rashba SOC as the only source of anisotropy (red), both Rashba and Dresselhaus linear terms SOC (blue) and 
cubic terms (black). Green broken line: additional EY mechanism could be the responsible for the observed 
anisotropy. 
}
\label{figres}
\end{figure*}

Now we focus on the six samples presented by Truitt {\it et al.} \cite{truitt}. % using Eq. (\ref{eq:ttwonorm2}). 
The experimental data are the dots of Fig. \ref{figres}, which we next attempt to fit using Eq. (\ref{eq:ttwonorm2}) (solid lines). 
%the angular dependency of $T_2^{-1}(\theta)$ for six different samples 
%(square points in Fig. \ref{figres}). 
%In our approach, each  sample is characterized by the (known) scattering time,  $\tau_k$  and the relative SOC 
%coupling strengths, $\chi, \Gamma_{\alpha,\beta}$, which are sample-dependent parameters \cite{prada_njop}.
%the two-parameter $\chi$, $\varphi$ ($\beta_3$, $\varphi$) linear (cubic) SOC anisotropy model.
%The results are summarized in Fig {\ref{figres}}, where the dots correspond to the experimental data. 
Once again, a fit that includes only one linear term for the SOC, $\chi=0$ (i. e., only one source of linear SOC), 
failed to reproduce %re unsuccessfully fitted considering linear Rashba as the only SOC source 
most of the experimental results (red curves), as it was attempted by Truitt {\it et al.}.
Only the narrower sample (IBM-01) agreed to this DP, linear-in momentum fit. 
This is consistent with the established picture where SOC strengths are sample dependent and 
very sensitive to well widths \cite{prada_njop, kohda}.
The blue lines correspond to the up-bound limit of anisotropy by including 
both Rashba and Dresselhaus SOC, which is still far from reasonable. 
Including both linear terms and cubic (black curves) contributions of SOC allows to  reproduce most 
of the experimental data. %We have assumed  $\varphi = \pi/4$. 
%The importance of the cubic term is apparent. 
However, we need make a distinction between the left and right panels of Fig. \ref{figres}: 
The former present larger mobility and a cubic fit seem to agree very well with the data, as those in Fig. \ref{figres2},  whereas the latter ones have a lower mobility 
and fail to reproduce DP relaxation, especially for sample `UW-31203'.  

In the low mobility limit,  %(or more frequent impurity scattering), 
tilted fields  enhance surface roughness scattering, resulting from the squeezing 
of the 2D well in the growth direction \cite{syphers_88}. 
Scattering processes  would then lead to anisotropic Elliot-Yafet (EY) mechanism for relaxation \cite{averkiev}, 
which is proportional to the scattering time, $T_2^{-1} = \alpha_{\rm EY}(\theta)\tau_k^{-1}$. 
Here the EY coefficient 
$\alpha_{\rm EY}$ depends on the amount of admixture of different spin states and reflects the probability that 
a spin-flip process occurs in a momentum scattering event. 
This effect has been already experimentally observed in the lower mobility samples \cite{graeff,wilamowski,studer}. 
A linear dependence of dephasing on a parallel magnetic field has indeed been attributed to an inhomogeneous dephasing 
in the EY limit \cite{studer, kikkawa} due to a variation $\Delta g$ of the $g$-factor,  
$\alpha_{EY}\sim \Delta g \mu_B B\sin{\theta}/2\hbar$. 
We identify this source with our discrepancy with the theoretical fit 
on `UW-21203'. The inset of the upper right panel of Fig. \ref{fig2} represents the discrepancy of 
the DP fit with the experimental data, showing a linear dependency for low tilted angles, 
which is well fitted with %one parameter, 
$\Delta g \simeq 10^{-4}$. This is  in excellent agreement with the experimental value \cite{graeff,wilamowski_prb66}. 
It is worth noting that  `UW-31203' data presents a pronounced `cusp-like' behavior (see upper-right panel of Fig. \ref{figres}), 
enforcing our arguments. The same argumentation may be applied to samples uw-31124 and uw-31121, where 
a cusp appears evident. Again, $\varphi$-resolved measurements would yield the answer.

{\it Note added}: During the final stage of this manuscript preparation, Song {\it et al.} 
reported on  EY-mechanism in Si QWs. This mechanism is enhanced in lower mobility samples \cite{song}. Their research 
enforces the last argument presented for the lower mobility samples of Fig. \ref{figres}. 

%We now consider other existing experimental results: 

%(only sample `IBM-01' had a different width, of 8 nm, see \cite{truitt}).

%The key point is that the scattering time is small

\section{Conclusions} 
\label{sec:conclusion}
In summary, we derived an expression for the angular dependency of the decoherence time in a quantum well, 
where the choice of the appropriate coordinate system is crucial. 
We find that %did a theoretical study on the anisotropy of the decoherence in typical quantum wells. 
the interplay of the Rashba and Dresselhaus SOC is reflected in the anisotropy of $T_2$,  
suggesting a practical scheme to determine their relative strength in the linear-in-momentum limit.
Although frequently ignored, we predict strongly enhanced DP processes due to the cubic terms of the SOC.
Experimental data owing to high mobility samples agree very well with our theoretical assumptions. 
Additional Elliot-Yafet mechanisms  may explain further discrepancies in samples with lower mobility.

{\it Acknowledgments.}  
We are grateful to A. Chudnovskiy and R. Joynt for enlightening discussions.
This work was supported by the Deutsche Forschungsgemeinschaft via the Graduiertenkolleg 1286 
"Functional Metal-Semiconductor Hybrid Systems".

\appendix 
\section{SIA and BIA in magnetic coordinates}
\label{app}
In order to express the Hamiltonian of Eq. (\ref{eq_h}) in the magnetic coordinates, 
we employ the Euler-Rodrigues formulation, for which the rotation matrix of angle $\theta$ around 
an axis defined by $\hat n = -\hat i \sin{\varphi} + \hat j\cos{\varphi}$ (see Fig. \ref{fig1}) is given by:  
\begin{widetext}
\begin{equation*}
   R^\theta_{\hat n}=\left(
        \begin{array}{ccc}
                1-\cos^2\varphi(1-\cos\theta) & 
                -\cos\varphi\sin\varphi(1-\cos\theta)
                & \cos\varphi\sin\theta \\
                -\cos\varphi\sin\varphi(1-\cos\theta)& 
                1-\sin^2\varphi(1-\cos\theta) & 
                \sin\varphi\sin\theta \\
                -\cos\varphi\sin\theta & 
                -\sin\varphi\sin\theta& 
                \cos\theta 
        \end{array}
   \right) \equiv \left(
        \begin{array}{ccc}
                F_1 & 
                -F_3
                & F_4 \\
                -F_3& 
                F_2& 
                F_5\\
                -F_4 & 
                -F_5& 
                F_6
        \end{array}
   \right).
\end{equation*}
\end{widetext}
Without loosing generality, we focus on samples grown along $\hat k$ = [001] direction, 
and set the other main crystallographic directions as $\hat i \parallel [100]$, $\hat j \parallel [010]$. 
In magnetic axes, we thus have: 
\begin{eqnarray}
\hat i_{x,y,z}& = & {R}(\theta,\hat n^\prime)\hat i = (F_1, -F_3, -F_4);\nonumber \\
\hat j_{x,y,z}& = & {R}(\theta,\hat n^\prime)\hat j = (-F_3, F_2, -F_5);\nonumber\\
\hat k_{x,y,z}& = & {R}(\theta,\hat n^\prime)\hat k = (F_4, F_5, F_6). \label{eqitox}
\end{eqnarray}
On the other hand, we have that the direction of $\hat k$ changes in time, and hence we define 
the angle $\phi_k(t)$:
\begin{eqnarray}
\vec k & =& %(k_x, k_y, k_z) = 
k_F(\hat i \cos{\phi_k} +\hat j \sin{\phi_k}) = 
k_F\left[
(F_1\cos{\phi_k}- F_3\sin{\phi_k})\hat x +
\right. \nonumber \\ &+&\left.
(-F_3\cos{\phi_k}+ F_2\sin{\phi_k})\hat y -
(F_4\cos{\phi_k}+ F_5\sin{\phi_k})\hat z 
\right].\nonumber\\
\label{kinmc}
\end{eqnarray}
Using Eq. (\ref{eqitox}) and (\ref{kinmc}) into the first term of Eq. (\ref{eq_h}), we obtain the expression of the 
Rashba Hamiltonian in magnetic coordinates:
\begin{eqnarray}
%\alpha( \vec\sigma\times\vec k)\cdot \hat{k} 
H_R
&=  \alpha k_F\left\{ \right. & \left.
\sigma_x[(F_4F_5-F_3F_6)\cos{\phi_k} + (F_6F_2+F_5^2)\sin{\phi_k} ]-
\right. \nonumber \\ &&\left. 
\sigma_y[(F_4^2+F_1F_6)\cos{\phi_k} + (F_4F_5-F_6F_3)\sin{\phi_k} ]+
\right. \nonumber \\ &&\left. 
\sigma_z[(F_1F_5+F_3F_4)\cos{\phi_k} - (F_3F_5+F_2F_4)\sin{\phi_k} ]
\right\}.\nonumber\\ %\equiv 
\label{eq:Rashba}
\end{eqnarray}
%Whereas for $\theta\simeq \pi/2$, $\varphi_k$ changes randomly in intervals given by $\tau_k$, 
For small $\theta$ we have $\phi_k(t) \simeq \omega_c t$ and $\omega_c \simeq (eB/m^*c)\cos{\theta}$ being the cyclotron frequency, 
and we obtain: 
\begin{equation}
H_{\rm R} = \vec{\sigma} \vec B_R^{\mathrm{eff}}, \quad
\vec B_R^{\mathrm{eff}} = \alpha k_F(\hat i \sin{\omega_ct}-\hat j \cos{\omega_ct}) = 
{\rm Re}\left\{ e^{i\omega_ct}\vec Z_{R} \right\}, 
\label{eq:RashbaB}
\end{equation}
which is an effective field that rotates in the plane with direction $\omega_c\hat k$.
%$ \vec Z_{R}(\theta=0) = \alpha k_F (1, -i)$
Expressed in the magnetic axis, we may use (\ref{eq:Rashba}) to get:
\begin{equation}
\vec Z_R(\theta,\varphi) = \alpha k_F\sum_q\left(f_{qR}^c(\theta,\varphi) -i f_{qR}^s(\theta,\varphi)\right)\hat q, \quad q=x,y,z
\label{eq:zr}
\end{equation}
with
\begin{eqnarray}
f_{xR}^c =& \cos{\varphi}\sin{\varphi}(1-\cos{\theta}); \quad  
&f_{xR}^s = \sin^2{\varphi} +  \cos^2{\varphi}\cos{\theta} ; \nonumber \\ 
f_{yR}^c =&  \cos^2{\varphi} +  \sin^2{\varphi}\cos{\theta} ;   \quad  
&f_{yR}^s = \cos{\varphi}\sin{\varphi}(1-\cos{\theta}); \nonumber \\ 
f_{zR}^c =& \sin{\varphi} \sin{\theta}; \quad 
&f_{zR}^s = -\sin{\varphi} \sin{\theta}. 
%\label{eq:fqr}
\end{eqnarray}

We consider next the Dresselhaus term for the lowest subband, assuming zero average momentum along the growth direction,
$\langle \vec k \cdot \hat k\rangle=0$: % \ Hence, we have:
\[
H_D = \beta k_F (\hat\sigma_ik_i - \hat\sigma_jk_j) =
\beta k_F (\hat\sigma_i \cos{\phi_k} -\hat\sigma_j \sin{\phi_k}),
\]
where we have defined $\beta=\gamma_{\rm D} \langle(k^2_j-k^2_k)\rangle$, and we have set:
$\langle k^2_j\rangle=\langle k^2_i\rangle$.
As before, we express the Dresselhaus Hamiltonian in the magnetic axes, for which we transform
 $\hat\sigma_{i,j}$  to get the rotated $\hat\sigma_{x,y,z}$:
\begin{eqnarray}
\hat\sigma_{i} &\to & {R}(\theta,\hat n^\prime)\hat\sigma_{i} = F_1\hat\sigma_x-F_3\hat\sigma_y - F_4\hat\sigma_z\nonumber\\
\hat\sigma_{j} &\to & {R}(\theta,\hat n^\prime)\hat\sigma_{j} = -F_3\hat\sigma_x+F_2\hat\sigma_y - F_5\hat\sigma_z,\nonumber\end{eqnarray}
yielding: 
%so we can express the Dresselhaus term as well in the rotated magnetic field as:
\begin{eqnarray}
H_D &= \beta k_F [
\hat\sigma_x (F_1\cos{\phi_k} +F_3\sin{\phi_k} )-
\hat\sigma_y (F_3\cos{\phi_k} +F_2\sin{\phi_k} )+
\nonumber \\ &+
\hat\sigma_z (-F_4\cos{\phi_k} +F_5\sin{\phi_k} )]. 
\label{eq:Dress}
\end{eqnarray}
The direction of the associated effective field rotates anti-parallel to $B_0$,
\begin{equation}
H_{\rm D} = \vec{\sigma} \vec B_D^{\mathrm{eff}}, \quad
\vec B_D^{\mathrm{eff}} = \beta k_F(\hat i \cos{\omega_ct}-\hat j \sin{\omega_ct}) = 
{\rm Re}\left\{ e^{-i\omega_ct}\vec Z_{D} \right\}, 
\label{eq:DressB}
\end{equation}
which is an effective field that rotates in the plane with direction $-\omega_c\hat k$, hence 
in opposite direction  to the Rashba field. 
In terms of the magnetic axis, we use (\ref{eq:Dress}) to get:
\begin{equation}
\vec Z_D(\theta,\varphi) = \beta k_F(f_{qD}^c -i f_{qD}^s)\hat q, \quad q=x,y,z
\label{eq:zd}
\end{equation}
with
\begin{eqnarray}
f_{xD}^c &= 1-\cos^2\varphi(1-\cos\theta) ; \quad 
&f_{xD}^s = \cos\varphi\sin\varphi(1-\cos\theta); \nonumber \\ 
f_{yD}^c &= -\cos\varphi\sin\varphi(1-\cos\theta); \quad 
&f_{yD}^s = \sin^2\varphi(1-\cos\theta) - 1; \nonumber \\ 
f_{zD}^c &= -\cos{\varphi}\sin{\theta}; \quad 
&f_{zD}^s = \sin{\varphi}\sin{\theta}. 
\label{eq:fqd}
\end{eqnarray}

As we can see $H^\prime$ has  terms in $\sigma_z$,  which are proportional to $\sin{\theta}$ %($F_{4,5}\propto \sin{\theta}$), 
and contribute to both decoherence and relaxation processes. 

SIA and BIA terms of $n$-order in momentum may be obtained using this procedure, only that the `effective magnetic fields' 
rotate in time at a higher frequency, which is  $n$-times faster than those for the linear terms. 
% with $n$ corresponding to the order of the momentum. 
For the cubic terms, we have: 
\begin{eqnarray}
\vec B_{3R}^{\mathrm{eff}} = \alpha_{3}(\vec k^3) (\hat i \sin{(3\omega_ct)}-\hat j \cos{(3\omega_ct)}),\nonumber \\
\vec B_{3D}^{\mathrm{eff}} = \beta_{3}(\vec k^3) (\hat i \cos{(3\omega_ct)}-\hat j \sin{(3\omega_ct)}). 
        %= \beta k_F \vec\omega_D^{\mathrm{eff}},\quad \omega_R^{\mathrm{eff}}= 
        %-\omega_c\hat k
\end{eqnarray}
yielding the components listed in table \ref{table1}.

\section{Bloch equations}
\label{app2}
To evaluate the trace of the double commutator on second term of Eq. (\ref{eq:ni2}),
we sum over all possible states and transform back to Schr\"odinger representation using Eq (8).
Adopting the summation convention,
%\begin{widetext}
\begin{eqnarray}
&{\rm Tr} &\left\{
\int_0^\infty \right. \left.
\left[[\sigma_i^*(t),\sigma^*_{q}(t-\tau)],\sigma_{q^\prime}^*(t)\right]\sigma_r^*(t)
u_{qq^\prime}(\tau)\ud \tau\right\} =  
%\overline{B_q^{\rm{eff}}(t-\tau)B_{q^\prime}^{\rm{eff}} (t) } \ud \tau\right\}
%\right\}= 
%\nonumber \\ 
\nonumber \\ && = 
n_i(t)  
%\int_0^\infty %\right.&\left.
\langle \alpha |\sigma_i| \beta\rangle
\langle \beta |\sigma_q |\gamma\rangle
\langle \gamma |\sigma_{q^\prime} |\delta\rangle
\langle \delta |\sigma_r |\alpha\rangle %e^{-i\omega_{\beta\gamma}\tau}
b_{qq^\prime}(\omega_{\beta\gamma}) + 
%u_{qq^\prime}(\tau) \ud \tau
\nonumber \\ &&+ 
n_i(t)  
%\int_0^\infty %\right.&\left.
\langle \alpha |\sigma_{q^\prime} |\beta\rangle
\langle \beta |\sigma_q |\gamma\rangle
\langle \gamma |\sigma_i |\delta\rangle
\langle \delta |\sigma_r |\alpha\rangle %e^{-i\omega_{\beta\gamma}\tau}
b_{qq^\prime}(\omega_{\beta\gamma}) -
\nonumber \\ && 
- n_i(t)  
%\int_0^\infty %\right.&\left.
\langle \alpha |\sigma_q |\beta\rangle
\langle \beta |\sigma_i |\gamma\rangle
\langle \gamma |\sigma_{q^\prime} |\delta\rangle
\langle \delta |\sigma_r |\alpha\rangle %e^{-i\omega_{\beta\gamma}\tau}
b_{qq^\prime}(\omega_{\alpha\beta})-
\nonumber \\ &&- 
n_i(t)  
%\int_0^\infty %\right.&\left.
\langle \alpha |\sigma_{q^\prime} |\beta\rangle
\langle \beta |\sigma_i |\gamma\rangle
\langle \gamma |\sigma_{q} |\delta\rangle
\langle \delta |\sigma_r |\alpha\rangle %e^{-i\omega_{\beta\gamma}\tau}
b_{qq^\prime}(\omega_{\gamma\delta}) = 
\nonumber \\ && =
%n_i(t) \left\{ b_{qq^\prime}(\omega_{\beta\gamma}) \left [ \langle \beta |\sigma_q |\gamma\rangle 
%\langle \gamma |\sigma_i\sigma_{q^\prime}\sigma_r |\beta\rangle +
%%\right.\right. \nonumber \\ &&\left. \left.
%\langle \beta |\sigma_q |\gamma\rangle  
%\langle \gamma |\sigma_i\sigma_r \sigma_{q^\prime} |\beta\rangle\right]-
%%\right.\right. \nonumber \\ &&\left. \left.
%\right. \nonumber \\ &&\left. 
%b_{qq^\prime}(\omega_{\alpha\beta}) \langle \alpha |\sigma_q |\beta\rangle 
%\langle \beta |\sigma_i\sigma_{q^\prime}\sigma_r |\alpha\rangle  -
%\right. \nonumber \\ &&\left. 
%-b_{qq^\prime}(\omega_{\delta\gamma})
%\langle \gamma |\sigma_q |\delta\rangle 
%\langle \delta |\sigma_r\sigma_{q^\prime} \sigma_r |\gamma\rangle \right\}= 
%\nonumber \\ &=&
%n_i(t)b_{qq^\prime}(\omega_{\alpha\beta}) \langle \alpha |\sigma_q |\beta\rangle 
%\langle \beta |[\sigma_i\sigma_{q^\prime}\sigma_r
%+\sigma_i\sigma_r \sigma_{q^\prime} 
%-\sigma_i\sigma_{q^\prime}\sigma_r 
%-\sigma_r\sigma_{q^\prime} \sigma_r
%|\alpha\rangle =\nonumber \\ &=&
n_i(t)b_{qq^\prime}(\omega_{\alpha\beta})
\langle \alpha |\sigma_q |\beta\rangle \langle \beta |
\left[[\sigma_{q^\prime},\sigma_r],\sigma_i\right]| \alpha\rangle
\end{eqnarray}
%\end{widetext}
Using the identities (\ref{eq:ids}):
\[
\left[[\sigma_{q^\prime},\sigma_r],\sigma_i\right] = 
(2i)^2 \varepsilon_{q^{\prime} r l }\varepsilon_{l i k } 
= -4[\delta_{q^{\prime} i} \delta_{r k} - \delta_{q^{\prime} k} \delta_{r i} ],
\]
Eq. (\ref{eqRed}) follows straightforward.
\newline
We now evaluate Eq. (\ref{eqRed}) for all three possible values of $r$.  First, for $r=x$, we have:
\begin{equation}
\label{eqnx}
\frac{\ud n_x(t)}{\ud t} =% \frac{2B_0}{\hbar} \epsilon_{zyx} n_y(t) +
\frac{2}{\hbar^2} n_i(t) (\delta_{q^\prime i}\delta_{xk}-\delta_{q^\prime k}\delta_{i x} )
\langle \alpha | \sigma_q |\beta \rangle \langle \beta | \sigma_k |\alpha \rangle 
%e^{-i\omega_{\alpha\beta}\tau} 
b_{qq^\prime}(\omega_{\alpha\beta}). 
\end{equation}
The Pauli matrices for spin $1/2$ read,
\[
\sigma_x = \left(
\begin{array}{cc} 0& 1\\1&0\end{array}
\right);
\quad\sigma_y = \left(
\begin{array}{cc} 0& -i\\i&0\end{array}
\right);
\quad\sigma_z = \left(
\begin{array}{cc} 1& 0\\0&-1\end{array}
\right);
\]
and  $|\alpha\rangle = \uparrow,\downarrow$ and $|\beta\rangle = \uparrow,\downarrow$.
For the second factor on the right, there are two terms. The first one, has $k=x$, so
$q^\prime = x,y,z$ and $q = x,y$, giving:
\begin{eqnarray}
&&\frac{2}{\hbar^2} n_{q^\prime}(t) 
\langle \alpha | \sigma_q |\beta \rangle \langle \beta | \sigma_x |\alpha \rangle 
b_{qq^\prime}(\omega_{\alpha\beta}) =  
\nonumber \\ &=&
\frac{2}{\hbar^2} \left\{
n_x[ b_{xx}(\omega_l) + b_{xx}(-\omega_l) 
- i b_{yx}(\omega_l) +i b_{yx}(-\omega_l)] + \right. \nonumber \\ &&\left. +
n_y[ b_{xy}(\omega_l) + b_{xy}(-\omega_l) -  ib_{yy}(\omega_l) + i b_{yy}(-\omega_l)] +
\right. \nonumber \\ &&\left. +
n_z   [ b_{xz}(\omega_l) + b_{xz}(-\omega_l) - ib_{yz}((\omega_l) +i b_{yz}(-\omega_l)]\right\}, 
\nonumber
\end{eqnarray}
and for the second term, $i=x$ while $k=q^\prime$, giving:
\begin{eqnarray}
&-&\frac{2}{\hbar^2} n_{x}(t) 
\langle \alpha | \sigma_q |\beta \rangle \langle \beta | \sigma_{q^\prime} |\alpha \rangle 
b_{qq^\prime}(\omega_{\alpha\beta}) =
\nonumber \\ &&
 -  \frac{2}{\hbar^2} \left\{
n_x[ b_{xx}(\omega_l) + b_{xx}(-\omega_l) + i b_{xy}(\omega_l) -i b_{xy}(-\omega_l) - \right. \nonumber \\ &&\left.
- ib_{yx}(\omega_l) +i b_{yx}(-\omega_l) + b_{yy}(\omega_l) + b_{yy}(-\omega_l) + 
 2b_{zz}(0) ]\right\}.  
\nonumber
\end{eqnarray}
On what follows, we will assume $b_{qq}(\omega) = b_{qq}(-\omega)$. %and $b_{qq^\prime} =0$ for $q\neq q^\prime$. 
Inserting these last two equations into (\ref{eqnx}), we get:
\begin{eqnarray}
\label{eqnx2}
\frac{\ud n_x(t)}{\ud t} &=& %\frac{-2B_0}{\hbar}n_y(t) 
-\frac{4}{\hbar^2} \left\{
n_x [ b_{yy}(\omega_l) + %\right.\nonumber \\ &+&\left. i b_{xy}(\omega_l) -i b_{xy}(-\omega_l) 
 b_{zz}(0)] -  
%-\right.\nonumber \\ &&-\left.
n_y [ b_{xy}(\omega_l) ] 
-\right.\nonumber \\ &&-\left.
n_z   [ b_{xz}(\omega_l)  ] % - ib_{yz}((\omega_l) +i b_{yz}(-\omega_l)]  
\right\}.
\end{eqnarray}
We now evaluate Eq. (\ref{eqRed}) for $r=y$,
\begin{equation}
%\label{eqny}
\frac{\ud n_y(t)}{\ud t} = %\frac{2B_0}{\hbar} \epsilon_{zxy} n_x(t) +
\frac{2}{\hbar^2} n_i(t) (\delta_{q^\prime i}\delta_{yk}-\delta_{q^\prime k}\delta_{i y} )
\langle \alpha | \sigma_q |\beta \rangle \langle \beta | \sigma_k |\alpha \rangle 
%e^{-i\omega_{\alpha\beta}\tau} 
b_{qq^\prime}(\omega_{\alpha\beta}). 
\nonumber
\end{equation}
For the second factor on the right, there are two terms. The first one, has $k=y$, so
$q^\prime = x,y,z$ and $q = x,y$, giving:
\begin{eqnarray}
&&\frac{2}{\hbar^2} n_{q^\prime}(t) 
\langle \alpha | \sigma_q |\beta \rangle \langle \beta | \sigma_y |\alpha \rangle 
b_{qq^\prime}(\omega_{\alpha\beta}) =  
\nonumber \\ &&
\frac{2}{\hbar^2} \left\{
n_x[ ib_{xx}(\omega_l) -i b_{xx}(-\omega_l)
+b_{yx}(\omega_l) + b_{yx}(-\omega_l)] + \right. \nonumber \\ &+&\left.
n_y[ ib_{xy}(\omega_l) - ib_{xy}(-\omega_l) + b_{yy}(\omega_l) + b_{yy}(-\omega_l)] +
\right. \nonumber \\ &+&\left.
n_z   [  ib_{xz}(\omega_l) - i b_{xz}(-\omega_l) + b_{yz}((\omega_l) + b_{yz}(-\omega_l)]\right\}, 
\nonumber
\end{eqnarray}
and for the second term, $i=y$ while $k=q^\prime$, giving:
\begin{eqnarray}
&-&\frac{2}{\hbar^2} n_{y}(t) 
\langle \alpha | \sigma_q |\beta \rangle \langle \beta | \sigma_{q^\prime} |\alpha \rangle 
b_{qq^\prime}(\omega_{\alpha\beta}) =
\nonumber \\ &-&  \frac{2}{\hbar^2} \left\{
n_y[ b_{xx}(\omega_l) + b_{xx}(-\omega_l) + i b_{xy}(\omega_l) -i b_{xy}(-\omega_l) - \right. \nonumber \\ &&\left.
- ib_{yx}(\omega_l) +i b_{yx}(-\omega_l) + b_{yy}(\omega_l) + b_{yy}(-\omega_l) + 
 2b_{zz}(0) ]\right\}.  
\nonumber
\end{eqnarray}
Collecting these last results, we get: 
%Inserting these last two equations into (\ref{eqny}), we get:
\begin{eqnarray}
\label{eqny2}
\frac{\ud n_y(t)}{\ud t} &=& % \frac{2B_0}{\hbar}n_x(t) +
\frac{4}{\hbar^2} \left\{
n_x [ b_{yx}(\omega_l) ] - 
%\right.\nonumber \\ &-&\left.
n_y [ b_{xx}(\omega_l)  + 2b_{zz}(0) ] +
\right.\nonumber \\ &+&\left.
n_z  [   b_{yz}((\omega_l)] 
\right\}.
\end{eqnarray}
Finally, for $r=z$ in (\ref{eqRed}), we get:
\begin{equation}
%\label{eqnz}
\frac{\ud n_z(t)}{\ud t} = %\frac{2B_0}{\hbar} \epsilon_{zxy} n_x(t) -
-\frac{2}{\hbar^2} n_i(t) (\delta_{q^\prime i}\delta_{zk}-\delta_{q^\prime k}\delta_{i z} )
\langle \alpha | \sigma_q |\beta \rangle \langle \beta | \sigma_k |\alpha \rangle 
%e^{-i\omega_{\alpha\beta}\tau} 
b_{qq^\prime}(\omega_{\alpha\beta}). \nonumber
\end{equation}
The first term on the right has $k=z$, and thus only $q=z$ contributes, with $q^\prime =i = x, y,z$, whereas
the second term on the right has $i=z$ with $q,q^\prime = x,y,z$,
\begin{eqnarray}
\frac{\ud n_z(t)}{\ud t} &=&\frac{4}{\hbar^2} \left\{ n_x(t)b_{zx}(0)   + n_y(t)  b_{zy}(0) +n_zb_{zz}(0)
\right.\nonumber \\&-&\left.
n_z(t)[ b_{xx}(\omega_l) + 
b_{yy}(\omega_l)   + 
%\right. \nonumber \\ &+&\left.
%ib_{xy}(\omega_l) -i b_{xy}(-\omega_l) -ib_{yx}(\omega_l) +i b_{yx}(-\omega_l) + 
b_{zz}(0)  ].
\right\}
\label{eqnzdt}
\end{eqnarray}
Equations (\ref{eqnx2},\ref{eqny2},\ref{eqnzdt}) are indeed the Bloch equations in (\ref{eq:Bloch}).

%%%%%%%%%%%%%%%%%%%%%%%%%%%%%%%%%%%%%%%%%%%%%%%%%%%%%%%%%%%%%

\begin{thebibliography}{99}

\bibitem{reviewSi} F. A. Zwanenburg, {\em et al.}, {\em Rev. Mod. Phys.} {\bf 85} 

\bibitem{SiComp}  J. L. Morton, D. R. McCamey, M. A. Eriksson, and S. A. Lyon. 
{\em Nature } {\bf 479}, 345--353 (2011).

\bibitem{rafa} R. Sanchez and G. Platero, {\em Phys. Rev. B} {\bf 87}, 081305(R) (2013).

\bibitem{datadas} S. Datta, and B. Das, {\em Appl. Phys. Lett.} {\bf 56}, 665 (1990).


\bibitem{zutic} I. \u Zuti\'c, J. Fabian, and S. Das Sarma, {\em Rev. Mod. Phys.} {\bf 76}, 323 (2004). 

\bibitem{tahan} C. Tahan, and R. Joynt, {\em Phys. Rev. B} {\bf 89} 075302 (2014).
\bibitem{kawakami} E. Kawakami, {\it et al.}, {\em Nature Nanotech.} {\bf 9} 666 (2014). 

\bibitem{tyryshkin_prl} %A. M. Tyryshkin {\em et al.}, {\em Physica E} {\bf 35}, 257--263 (2006); 
A. M. Tyryshkin, S. A. Lyon, W. Jantsch, and F. Sch{\"{a}}ffler, {\em Phys. Rev. Lett.} {\bf 94}, 126802 (2005); 
\bibitem{tyryshkin} %A. M. Tyryshkin {\em et al.}, {\em Physica E} {\bf 35}, 257--263 (2006); 
S. Shankar, A. M. Tyryshkin, J. He, and S. A. Lyon, {\em Phys. Rev. B} {\bf 82}, 195323 (2010); 
A. Morello {\em et al.}, {\em Nature} {\bf 467}, 687 (2010).

\bibitem{wong} C. H. Wong, M. A. Eriksson, S. N. Coppersmith, and M. Friesen, {\em Phys. Rev. B} {\bf 92}, 045403 (2015).
\bibitem{rohling} N. Rohling, M. Russ, and G. Burkard, {\em Phys. Rev. Lett.} {\bf 113}, 176801 (2014). 
\bibitem{veldhorst}M. Veldhorst {\em et al.}, {\em Nature} {\bf 526}, 410 (2015)

\bibitem{loss} D. Loss, and D. P. DiVincenzo, {\em Phys. Rev. A} {\bf 57}, 120--126 (1998); 
R. De Sousa, and S. Das Sarma, {\em Phys. Rev. B} {\bf 67}, 033301 (2003). 

\bibitem{sanada} H. Sanada, Y. Kunihashi, H. Gotoh, K. Onomitsu, M. Kohda, J. Nitta, P. V. Santos, and T. Sogawa, 
{\em Nature Phys.} {\bf 9}, 280--283 (2013).  
\bibitem{koppens} F. H. L. Koppens {\it et al.}, {\em Nature} {\bf 442}, 766 (2006). 

\bibitem{wrinkler} 
G. Dresselhaus, 
%Spin-orbit coupling effects in zinc blende structures. 
{\it Phys. Rev.} {\bf 100}, 580–586 (1955); 
R.~Wrinkler,
\newblock {\em Spin-Orbit Coupling Effects in Two-Dimensional Electron and Hole System}.
\newblock Springer--Verlag, Berlin--Heidelberg--New York, 2003. 
%\bibitem{dresselhaus} 
\bibitem{sherman} I. V. Tokatly, and E. Ya. Sherman, {\em Phys. Rev. B} {\bf 82}, 161305(R) (2010); 
M. M. Glazov, E. Ya. Sherman, and V. K. Dugaev, 
{\bf 42} 2157 (2010). 
\bibitem {vervoort}
L. Vervoort and P. Voisin, {\em Phys. Rev. B} {\bf 56}, 12744 (1997).

\bibitem{lyanda} Y. B. Lyanda-Geller, adn A. D. Mirlin, 
{\em Phys.\ Rev.\ Lett.} {\bf 72}, 1894 (1994).
\bibitem{prada_njop} M. Prada, G. Klimeck, R. Joynt, {\em New J. Phys.} {\bf 13}, 013009 (2011). 
\bibitem{nestoklon} M. O. Nestoklon, L. E. Golub, E. L. Ivchenko, {\em Phys. Rev. B} {\bf 73}, 235334 (2006).
\bibitem{ganichev2} S. D. Ganichev, and L. E. Golub, {\em Phys. Stat. Sol. B}, {\bf 251} 1801 (2014).
\bibitem{cubic}
X. Cartoix\`a, L.-W. Wang, D.Z.-Y. Ting, and Y.-C. Chang,
{\em Phys.\ Rev. B} {\bf 73}, 205341 (2006); 
W. Yang and K. Chang,
{\em Phys.\ Rev.\ B} {\bf 74}, 193314 (2006); 
H. Nakamura, T. Koga and T. Kimura,
{\em Phys.\ Rev.\ Lett.} {\bf 108}, 206601(2012); 
R. Moriya {\em et al.}, 
{\em Phys.\ Rev.\ Lett.} {\bf 113}, 086601 (2014).
 
\bibitem{schli} C. P. Schlister, 
{\it Principles of Magnetic Resonance},  Springer Series in Solid-State Sciences, New York, (1963).  

\bibitem{abragam} A. Abragam, 
{\it The Principles of Nuclear Magnetism}, Oxford University Press, (2002). 

\bibitem{dp} M.~I. D'Yakonov and V.~I. Perel'.
\newblock {\em Sov.\ Phys.\ Solid State} 13:3023, 1971.

\bibitem{li_prl} F. Li, Y. V. Pershin, V. A. Slipko, and N. A. Sinitsyn, 
\newblock {\em Phys.\ Rev.\ Lett.} {\bf 111}, 067201 (2013).

\bibitem{redfield} A. G. Redfield, 
{\it IBM J. Res. Dev.} {\bf 1}, 19 (1957).
\bibitem{glazov} M. M. Glazov, 
{\em Phys.\ Rev. B} {\bf 70}, 195314 (2004).

\bibitem{kohda} M. Kohda {\it et al.}, {\em Phys. Rev. B} {\bf 86}, 081306(R) (2012). 
\bibitem{truitt}  Electron spin coherence in Si/SiGe quantum wells, 
J. L. Truitt, {\it et al.}, 
%K. A. Slinker, K. L. M. Lewis, D. E. Savage, C. Tahan, L. J. Klein, R. Joynt, M. G. Lagally, D. W. van der Weide, S. N. Coppersmith, M. Friesen and M. A. Eriksson, 
{\it cond-mat/0411735}, 
%"Electron spin resonance and related phenomena in low dimensional structures" 
Topics in Applied Physics Series {\bf 115}, ed. M. Fanciulli (Springer, 2008). 

\bibitem{syphers_88} D. A. Syphers, J. E. Furneaux, {\em Solid State Comm.} {\bf 65}, 1513--1515 (1988); 
T. Ando, A. B. Fowler, F. Stern, {\em Rev. Mod. Phys.} {\bf 54}, 437--672 (1982). 

\bibitem{averkiev} N. S. Averkiev, L. E. Golub, and M. Willander, {\em J. Phys: Condens. Matter} {\bf 14}, R271--R283 (2002); 
S. D. Ganichev {\em et al.}, {\em Phys. Rev. B} {\bf 68}, 081302(R) (2003). 

\bibitem{studer} M. Studer, S. Sch\"on, K. Ensslin, and G. Salis, {\em Phys. Rev. B} {\bf 79}, 045302 (2009). 

\bibitem{graeff} C. F. O. Graeff, M. S. Brandt, M. Stutzmann, M. Holzmann, G. Abstreiter, and F. Sch\"affler, 
{\em Phys. Rev. B} {\bf 59}, 13242 (1999). 

\bibitem{wilamowski} Z.~Wilamowski and W.~Jantsch.
\newblock {\em Phys.\ Rev.\ B} {\bf 69} 035328, 2004.

\bibitem{kikkawa} J. M. Kikkawa, and D. D. Awschalom, {\em Phys. Rev. Lett.} {\bf 80}, 4313 (1998). 

\bibitem{wilamowski_prb66} Z. Wilamowski, W. Jantsch, H. Malissa, and U. R\"ossler, 
{\em Phys. Rev. B} {\bf 66}, 195315 (2002). 

%\bibitem{elliot} R. J. Elliot, {\em Phys. Rev.} {\bf 96}, 266 (1954). 
\bibitem{jantsch_PE02} W. Jantsch, Z. Wilamowski, N. Sandersfeld, M. M\"uhlberger, and F. Sch\"affler,
{\em Physica E} {\bf 13}, 504 (2002).

%\bibitem{review} M. W. Wu, J. H. Jiang and M. Q. Weng, 
%{\em Physics Reports} {\bf 493}, 61–-236, (2010).  

%\bibitem{krich} J. J. Krich and B. I. Halperin, {\em Phys.\ Rev.\ Lett.} {\bf 98} 226802 (2007). 

%\bibitem{wilamowski_prl} Z.~Wilamowski, H. Malissa, F. Sch{\"{a}}ffler and W.~Jantsch.
%\newblock {\em Phys.\ Rev.\ Lett.} {\bf 98} 187203 (2007).  

%\bibitem{ensslin}L.~Meier, G.~Salis, I.~Shorubalko, E.~Gini, S.~Sch{\"{o}}n, and K.~Ensslin.
%\newblock {\em Nature Physics} 3:650, 2007.


\bibitem{song} Y. Song, adn S. Das Sarma, {\em	arXiv:1606.09578 [cond-mat.mes-hall]}
%Impurity-driven intervalley spin-flip scattering-induced 2D spin relaxation in silicon

\end{thebibliography}
\end{document}